\documentclass[10pt, article,twocolumn]{IEEEtran}
\usepackage{times}
\usepackage[cmex10]{amsmath}
\usepackage{amssymb}
\usepackage{epsfig,verbatim}
\usepackage{algorithm}
\usepackage{algpseudocode}

\usepackage{graphicx,color,epsfig,rotating,subfigure}
\usepackage{amsfonts,amsmath,amssymb}
\usepackage{algorithm}
\usepackage{subfigure}
\usepackage{algpseudocode}

\newcommand\blfootnote[1]{%
  \begingroup
  \renewcommand\thefootnote{}\footnote{#1}%
  \addtocounter{footnote}{-1}%
  \endgroup
}

\long\def\comment#1{}

\newfont{\bbb}{msbm10 scaled 700}

\newfont{\bb}{msbm10 scaled 1100}
\newcommand{\CC}{\mbox{\bb C}}
\newcommand{\PP}{\mbox{\bb P}}
\newcommand{\RR}{\mbox{\bb R}}

\newcommand{\EE}{\mbox{\bb E}}

\newcommand{\av}{{\bf a}}

\newcommand{\cv}{{\bf c}}

\newcommand{\ev}{{\bf e}}

\newcommand{\hv}{{\bf h}}

\newcommand{\rv}{{\bf r}}

\newcommand{\tv}{{\bf t}}

\newcommand{\wv}{{\bf w}}

\newcommand{\xv}{{\bf x}}
\newcommand{\yv}{{\bf y}}
\newcommand{\zv}{{\bf z}}

\newcommand{\Hm}{{\bf H}}

\newcommand{\Pm}{{\bf P}}

\newcommand{\Ac}{{\cal A}}
\newcommand{\Bc}{{\cal B}}
\newcommand{\Cc}{{\cal C}}
\newcommand{\Dc}{{\cal D}}
\newcommand{\Ec}{{\cal E}}

\newcommand{\Hc}{{\cal H}}
\newcommand{\Ic}{{\cal I}}

\newcommand{\Nc}{{\cal N}}

\newcommand{\Rc}{{\cal R}}
\newcommand{\Sc}{{\cal S}}

\newcommand{\Wc}{{\cal W}}

\newcommand{\SNR}{{\sf SNR}}

\newcommand{\eqdef}{\stackrel{\Delta}{=}}

\newcommand{\transp}{{\sf T}}

\newtheorem{theorem}{Theorem}

\newtheorem{definition}{Definition}

\newtheorem{lemma}{Lemma}

\newcommand{\argmax}{\operatornamewithlimits{argmax}}
\newcommand{\argmin}{\operatornamewithlimits{argmin}}

\title{Uplink Multiuser Massive MIMO Systems with Low-Resolution ADCs: \\A Coding-Theoretic Approach}

\author{
\IEEEauthorblockN{
              Song-Nam Hong\authorrefmark{1}, Seonho Kim\authorrefmark{1}, and Namyoon Lee\authorrefmark{2}}\\
\IEEEauthorblockA{\authorrefmark{1}Ajou University, Suwon, Korea,\\
              email: \{shkim1005, snhong\}@ajou.ac.kr}\\
\IEEEauthorblockA{\authorrefmark{2}POSTECH, Pohang, Korea,\\
              email: nylee@postech.ac.kr}
              \thanks{A part of this work was presented in \cite{Song_Kim_Lee17}.}
}

\begin{document}

\maketitle

\date{}

\blfootnote{
}

\begin{abstract}
This paper considers an uplink multiuser massive multiple-input multiple-output (MIMO) system with low-resolution analog-to-digital converters (ADCs), in which $K$ users with a single-antenna communicate with one base station (BS) with $N_{\rm r}$ antennas. In this system, we present a novel multiuser MIMO detection framework that is inspired by coding theory. The key idea of the proposed framework is to create a code $\Cc$ of length $2N_{\rm r}$ over a spatial domain. This code is constructed by a so-called {\em auto-encoding} function that is not designable but is completely described by a channel transformation followed by a quantization function of the ADCs. From this point of view, we convert a multiuser MIMO detection problem into an equivalent channel coding problem, in which a codeword of $\Cc$ corresponding to users' messages is sent over $2N_{\rm r}$ parallel channels, each with different channel reliability. To the resulting problem, we propose a novel {\em weighted} minimum distance decoding (wMDD) that appropriately exploits the unequal channel reliabilities. It is shown that the proposed wMDD yields a non-trivial gain over the conventional minimum distance decoding (MDD).  From coding-theoretic viewpoint, we 
identify that  bit-error-rate (BER) exponentially decreases with the minimum distance of the code $\Cc$, which plays a similar role with a condition number in conventional MIMO systems. Furthermore, we develop the communication method that uses the wMDD for practical scenarios where the BS has no knowledge of channel state information. Finally, numerical results are provided to verify the superiority of the proposed method.
\end{abstract}
\begin{keywords}
Massive MIMO, analog-to-digital converter (ADC), low-resolution ADC, multiuser MIMO detection.
\end{keywords}

\section{Introduction}


The use of a very large number of antennas at the base station (BS), referred to as massive multiple-input-multiple-output (MIMO), is one of the promising approaches to cope with the predicted wireless data traffic explosion \cite{Larsson, Lu}. The use of a large number of antennas at the BS can improve the capacity and energy efficiency  \cite{Larsson,Lu, Marzetta}. In contrast, it can considerably increase the hardware cost and the radio-frequency (RF) circuit consumption \cite{Yang}. Among all the components in a RF chain, a high-resolution analog-to-digital converter (ADC) is particularly power-hungry since the power consumption of an ADC is scaled exponentially with the number of quantization bits and linearly with the baseband bandwidth  \cite{Murmann, Mezghani-2011}. To overcome this challenge, the use of low-resolution ADCs  (e.g., 1$\sim$3 bits) for massive MIMO systems has received increasing attention over the past years. The one-bit ADC is particularly attractive due to the lower hardware complexity. In this case, the in-phase and quadrature components of the continuous-valued received signals are separately quantized using simple zero-threshold comparators; thereby, there is no need for an automatic gain controller \cite{Donnell, Hoyos}. Despite the benefits of using low-resolution ADCs, it gives rise to numerous technical challenges: 1) obtaining an accurate channel estimation at the receiver (CSIR) is complicated; 2) conventional MIMO detection methods, developed for linear MIMO systems, yield an unsatisfactory performance as it does not capture the impact of non-linearity of ADCs.

In recent, there have been extensive researches on the MIMO detection and channel estimation methods for uplink massive MIMO systems with low-resolution ADCs \cite{Mo2}-\cite{Studer}. Most of works have focused on the one-bit ADCs due to its simplicity and practical attractiveness. Numerous channel estimation methods were presented as least-square (LS) based method \cite{Risi}, maximum-likelihood (ML) type method \cite{Choi}, zero-forcing (ZF) type method \cite{Choi}, and Bussgang decomposition based method \cite{Li}. Regarding MIMO detection methods, the optimal ML detection was developed in \cite{Choi} and low-complexity methods were also proposed in \cite{Mollen, Mollen2}.  Beyond the one-bit ADCs, it is difficult to develop an optimal channel estimation and MIMO detection methods. Instead, several suboptimal methods were proposed as follows. A joint channel and data estimation method was developed in \cite{Li} using Bayesian inference theory. This method, however, is not practical due to its unaffordable computational complexity. In addition, the low-complexity methods as ZF MIMO detection \cite{Choi2} and minimum-mean-square-error (MMSE) MIMO detection  \cite{Mezghani2} were proposed. The major limitations of such methods are not satisfactory performance in particular when the number of receiver antennas is not so large \cite{Choi,Choi2,Mezghani2}.

\begin{figure}
\centerline{\includegraphics[width=9.5cm]{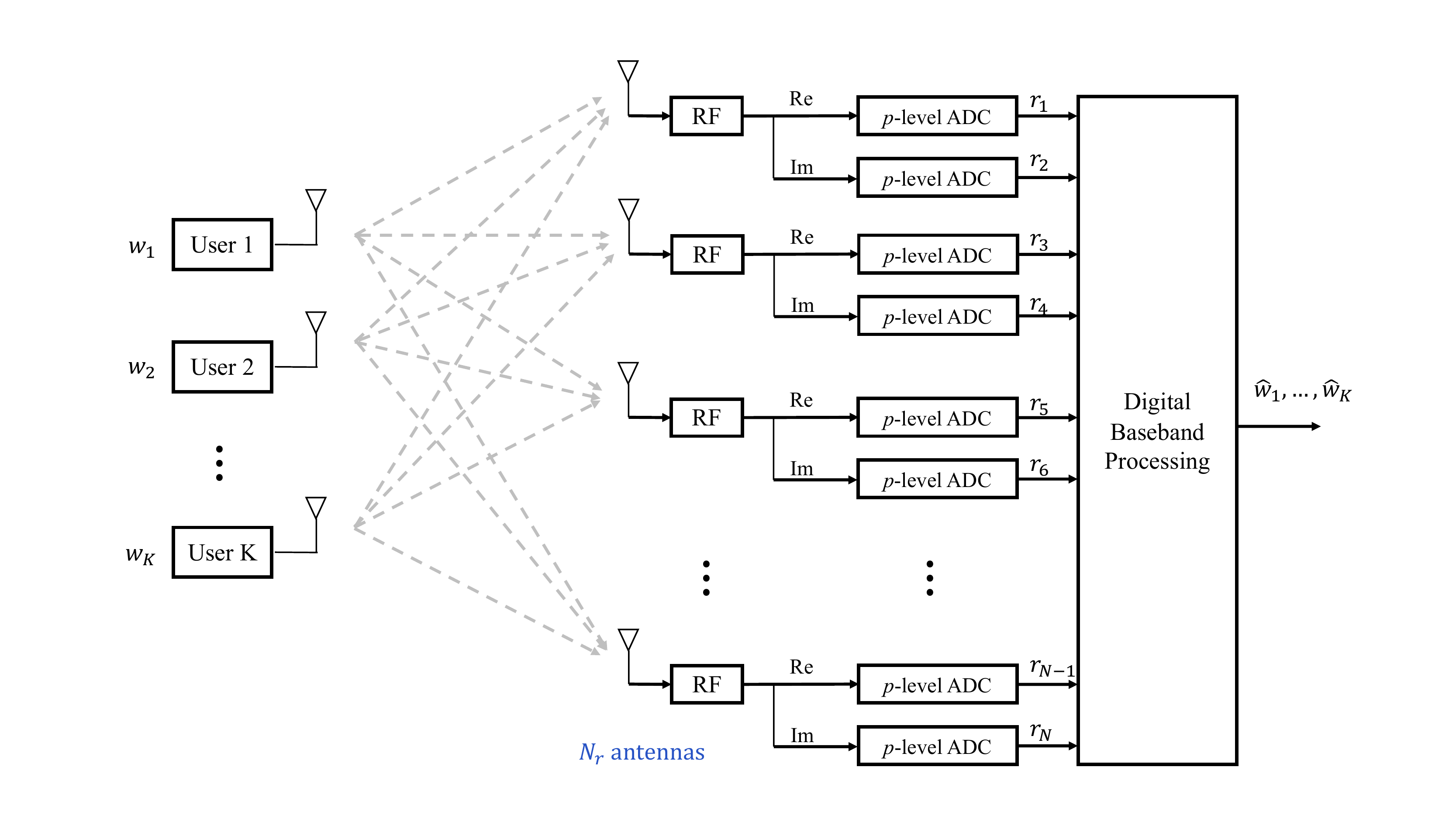}}
\caption{Uplink multiuser massive MIMO systems in which each receiver antenna at a BS is equipped with $p$-level ADC.}
\label{model}
\end{figure}

In this paper, we focus on the uplink multiuser massive MIMO systems in which  $K$ users equipped with a single-antenna communicate with one BS equipped with $N_{\rm r}$ antennas, as illustrated in Fig.~\ref{model}. Especially it is assumed that  each receiver is equipped with a RF chain followed by two $p$-level ADCs which are applied to each real and imaginary part separately. Each user transmits its signal from $m$-ary constellation set (e.g., QAM constellation). For such system, we present a novel multiuser MIMO detection framework inspired by coding theory.

The contributions of this paper are summarized as follows.

\begin{figure}
\centerline{\includegraphics[width=9cm]{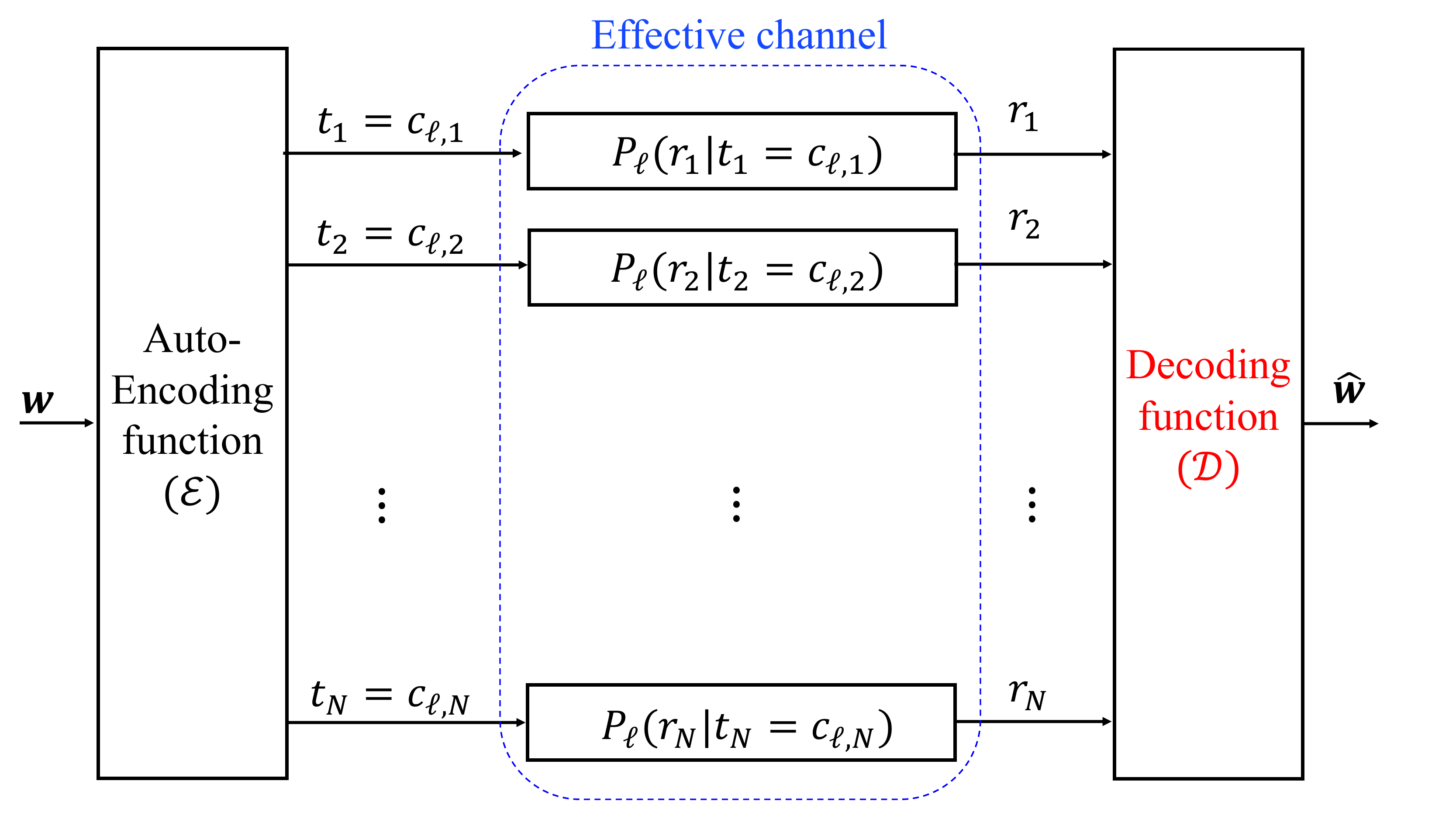}}
\caption{Illustration of an effective communication model to be used for the proposed coding method. Notice that an auto-encoding function $\Ec$ is determined as a function of $\Hm$ and a quantization function of ADCs and a decoding function $\Dc$ is proposed. In addition, the transition probabilities of the effective channel depend on message vector $\wv$ and channel matrix $\Hm$.}
\label{e_model}
\end{figure}

\begin{itemize}

\item Our major contribution is to present a novel multiuser MIMO detection framework by introducing an equivalent coding problem. The key idea of the proposed framework is to view a channel transformation followed by a quantization function of the ADCs (in short, a non-linear MIMO channel) as an auto-encoding function. This can create a code $\Cc$ (over a spatial domain) of length $2N_{\rm r}$, alphabet size $p$, and rate $\frac{K\log{m}}{2N_{\rm r}}$. As seen in Fig.~\ref{e_model}, a codeword of $\Cc$ is sent over $2N_{\rm r}$ parallel $p$-ary input/output channels, each with possibly different channel reliability. From this coding perspective, we identify that the minimum distance of the $\Cc$ plays a crucial role in determining the {\em goodness} of a channel matrix and show that bit-error-rate (BER) exponentially decays with the minimum distance.

\item In the equivalent coding problem, we present two decoding methods: i) minimum distance decoding (MDD) and ii) maximum likelihood decoding (MLD). To distinguish the MLD for an original MIMO channel in \cite{Choi}, the proposed MLD is referred to as eMLD. We compare the two decoding methods and identify their fundamental difference. Our crucial observation is that eMLD appropriately harnesses the unequal channel reliabilities of the resulting parallel channels. Because of this fact, eMLD can outperform MDD which does not exploit the channel reliability in decoding.

\item We present a novel {\em weighted} MDD (wMDD) as a practical approximation of eMLD. It follows the decoding procedures of MDD with the {\em weighted} Hamming distance, instead of Hamming distance used in MDD, where the weights are properly chosen so that a higher distance is assigned to a more reliable channel. This is reasonable as higher belief (i.e., higher weights) should be allocated to  more reliable subchannels as in the optimal eMLD, in which the optimality is with respect to the equivalent channel. It is also mathematically proved that wMDD outperforms MDD with the aid of weights.

\item Finally, we develop the communication method that implements the proposed wMDD for practical systems where a BS (or a receiver) has no knowledge of channel state information. Via simulations, we demonstrate that the proposed wMDD outperforms the existing MIMO detection methods.
\end{itemize}

The paper is organized as follows. In Section~\ref{sec:SM}, we describe the system model of uplink massive MIMO systems with $p$-level ADCs. Inspired by coding theory, in Section~\ref{sec:Main}, we present a novel multiuser MIMO detection framework. In Section~\ref{sec:WMD}, we propose a novel wMDD that appropriately exploits the unequal channel reliabilities of transformed parallel channels.
In Section~\ref{sec:practical}, we develop the communication framework based on the proposed wMDD and show that it can outperform the existing techniques for practical scenarios. Section~\ref{sec:conclusion} concludes the paper.

{\em Notation:} Lower and upper boldface letters represent column vectors and matrices, respectively. For any vector $\xv$, $d_{\rm w}(\xv)$ denotes the Hamming weight, i.e., the number of non-zero values in $\xv$. For any two vector $\xv$ and $\yv$, $d_{\rm h}(\xv,\yv)$ represents the Hamming distance, i.e., the number of positions at which the corresponding symbols are different. For any $k \in \{0,...,K-1\}$, we let $g(k)=[b_0,b_1,\ldots,b_{K-1}]^{\transp}$ represent the $m$-ary expansion of $k$ where 
$k=b_0m^0+\cdots+b_{K-1}m^{K-1}$ for $b_i \in \{0,...,m-1\}$.
 We also let $g^{-1}(\cdot)$ denote its inverse function. For a vector, $g(\cdot)$ is applied element-wise. Likewise, if a scalar function is applied to a vector, it will be performed element-wise. ${\rm Re}(\av)$ and ${\rm Im}(\av)$ represent the real and complex part of a complex vector $\av$, respectively.

\section{System Model}\label{sec:model}\label{sec:SM}

We consider a single-cell uplink system in which $K$ users with a single-antenna communicate with one  BS with an array of $N_{\rm r} > K$ antennas.  Let $w_k \in \mathcal{W}=\{0,...,m-1\}$ represent the user $k$'s message for $k \in\{1,...,K\}$, each of which contains $\log{m}$ information bits. We also denote $m$-ary constellation set by $\Sc=\{s_0,...,s_{m-1}\}$ with power constraint as
\begin{equation}
\frac{1}{m}\sum_{i=0}^{m-1}\|s_i\|^2 = \SNR.
\end{equation}
Then, the transmitted symbol of user $k$, ${\tilde x}_k$, is obtained by a modulation function $f:\mathcal{W}\rightarrow \Sc$ as
\begin{equation}
\tilde{x}_k = f(w_k) \in \Sc.
\end{equation} 
When the $K$ users transmit the symbols ${\tilde \xv}=[\tilde{x}_1,\ldots,\tilde{x}_K]^{\transp}$, the discrete-time complex-valued baseband received signal vector at the BS, ${\bf \tilde r}\in\mathbb{C}^{N_{\rm r}}$, is given by
\begin{equation}
{\bf \tilde r} = {\bf \bar H}{\bf \tilde x} +{\bf \tilde z}, \label{eq:system_complex}
\end{equation} where ${\bf \tilde H} \in \CC^{N_{\rm r} \times K}$ is the channel matrix between the BS and the $K$ users, i.e., the $i$-th row of ${\bf \tilde H}$ is the channel vector between the $i$-th receiver antenna at the BS and the $K$ users. In addition, ${\bf \tilde z}=[{\tilde z}_1,\ldots,{\tilde z}_{N_{\rm r}}]^{\transp}\in\mathbb{C}^{N_{\rm r}}$ is the noise vector whose elements are distributed as circularly symmetric complex Gaussian random variables with zero-mean and unit-variance, i.e., ${\tilde z}_i \sim \Cc\Nc(0,1)$.

In the MIMO system with low-resolution ADCs, each receiver antenna of the BS is equipped with a RF chain followed by two $p$-level ADCs that are separately applied to each real and imaginary part. Let $\phi_p(\cdot): \RR \rightarrow \{0,...,p-1\}$ represent the $p$-level ADC quantizer function with
\begin{equation}
\hat{r}=\phi_p(\tilde{r}).
\end{equation}  Then, the BS receives the quantized output vector as
\begin{align}
\hat{\rv}_{\rm R} = \phi_p({\rm Re}({\bf \tilde r}))\mbox{ and }\hat{\rv}_{\rm I} =\phi_p({\rm Im}({\bf \tilde r})).
\end{align} Although the proposed method can be applied to any ADC quantizer function, we in this paper restrict ourselves to a {\em stair-type} quantizer as
\begin{equation}
  \hat{r}  =\phi_p({\tilde r}) = \ell \mbox{ if }  {\tilde r} \in [\Delta_\ell, \Delta_{\ell-1}),\label{eq:p-ADC}
 \end{equation} for $\ell \in \{0,...,p-1\}$,  where $\Delta_{-1} = \infty$, $\Delta_{p-1} = - \infty$, and $\Delta_j> \Delta_k $ if $j < k$.

For the ease of representation, we rewrite the complex input-output relationship in \eqref{eq:system_complex} into the equivalent real representation as
\begin{equation}
\rv = \phi_p\left(\Hm\xv(\wv)+\zv\right), \label{eq:obs1}
\end{equation}
where $\rv=[\hat{\rv}_{\rm R}^{\transp},\hat{\rv}_{\rm I}^{\transp}]^{\transp}\in \{0,1,\ldots, p-1\}^{N}$, $\xv(\wv)=[\mbox{Re}(\tilde{\xv})^{\transp},\mbox{Im}(\tilde{\xv})^{\transp}]^{\transp}$, $\zv=[\mbox{Re}(\tilde{\zv})^{\transp},\mbox{Im}(\tilde{\zv})^{\transp}]^{\transp}\in\mathbb{R}^{N}$, and 
\begin{equation*}
\Hm = \left[ {\begin{array}{cc}
   \mbox{Re}({\bf \tilde H}) & -\mbox{Im}({\bf \tilde H}) \\      
   \mbox{Im}({\bf \tilde H}) & \mbox{Re}({\bf \tilde H})\\
 \end{array} } \right] \in\mathbb{R}^{N\times 2K},
 \end{equation*}
where $N=2N_{\rm r}$. This real system representation will be used in the sequel.

\section{The Proposed MIMO Detection Method}\label{sec:Main}

In this section, we present a novel multiuser MIMO detection method. From (\ref{eq:obs1}), the received signal of the $i$-th receiver antenna after the ADC quantizer is 
\begin{align}
r_i & =\phi_p\left(\hv^{\transp}_i \xv(\wv) + z_i\right) \mbox{ for }i\in\{1,...,N\}, \label{eq:o_channel}
\end{align} where $\wv=[w_1,\ldots,w_K]^{\transp}$ represents all the users' messages that creates channel input vector ${\bf x}({\bf w})$.

Our goal is to develop a multiuser MIMO detection method to estimate users' messages $\hat{\wv}$ from the observations $\rv=[r_1,\ldots,r_{N}]^{\transp}$.  Instead of directly solving the above problem, we will solve it by introducing an {\em equivalent communication model} from a coding theory perspective. 
As illustrated in Fig.~\ref{e_model}, the proposed model consists of three parts as
\begin{itemize}
\item {\bf Auto-encoding function}: This maps the users' messages $\wv=g(\ell) \in \{0,...,m-1\}^K$ into a $p$-ary codeword $\cv_{\ell} \in \Cc$ of length $N$ and rate $\frac{K}{N}\log{m}$. Notice that the code $\Cc$ is not designable but is completely characterized as a function of the non-linear MIMO channel;
\item {\bf Effective channel}: This is composed of $N$ parallel $p$-ary input/output channels with transition probabilities $P_{\ell}(r_i|t_i=c_{\ell,i})$ for $\ell \in \{0,...,m^K-1\}$ and $i \in \{1,...,N\}$. Notice that the transitional probabilities depend on input vectors;
\item {\bf Decoding function}: This maps the observation $\rv$ into users' message $\hat{\wv}$, which is what we will propose in this section, which is also the proposed multiuser MIMO detection method in the original channel.
\end{itemize}

\subsection{The Proposed Coding-Theoretic Framework}\label{subsec:PF}

We specify an encoding function, an effective channel, and a decoding function of the equivalent communication model as illustrated in Fig.~\ref{e_model}.

\subsubsection{Auto-encoding function} 

We define a code $\Cc$ over a spatial domain as
\begin{equation}
\Cc=\{\cv_0,\ldots,\cv_{m^K-1}\},
\end{equation} where each codeword $\cv_{\ell}$ is defined as
\begin{equation*}
\cv_\ell= \left[\phi_p\left(\hv_{1}^{\transp}\xv(g(\ell))\right),\ldots, \phi_p\left(\hv_{N}^{\transp}\xv(g(\ell))\right)\right]^{\transp}.
\end{equation*} The code $\Cc$ is a non-linear code of length $N$, alphabet size $p$, and code rate $\frac{K\log{m}}{N}$.

We refer to this code as a {\em channel-dependent} code, because it is completely characterized by channel matrix $\Hm=[\hv_1,\ldots,\hv_{N}]^{\transp}$ and quantization function $\phi_p(\cdot)$.  Letting $\Ic=\{0,...,m^K-1\}$ be the index set for the codewords of $\Cc$, there exists the one-to-one mapping $g:\Wc^{K} \rightarrow \Ic$ between the message of a user and a codeword index as $\wv = g(\ell)$ and $\ell   = g^{-1}(\wv)$. Furthermore, let $d_{\rm min}(\Hm)$ denote the minimum distance of the code $\Cc$ associated with a channel matrix $\Hm$, defined by
\begin{equation}
d_{\rm min}(\Hm) \eqdef \min_{i,j \in \Ic_{\Cc}: i \neq j} d_{\rm h}(\cv_i,\cv_j).
\end{equation} In classical coding theory, this parameter plays a fundamental role in determining the code performance, especially at high SNRs \cite{MacWilliams}. So, does it in our problem (see Section~\ref{subsec:DG} for details). 

In Fig.~\ref{e_model}, the input $\tv$ of an effective channel is generated by an auto-encoding function $\Ec :\{0,...,m-1\}^K \rightarrow \Cc$ as
\begin{equation}
\tv=\Ec(\wv) = \cv_{\ell}
\end{equation}  where $\ell = g^{-1}(\wv)$.\\

{\bf Example 1:} Consider a $2\times 2$ MIMO system with one-bit ADC, and each user is assumed to use QPSK modulation, i.e.,  $N_{\rm r}=2$, $K=2$, $p=2$, and $m=4$. Then, for a given channel matrix ${\bf H}\in\mathbb{R}^{4\times 4}$, one can create a code $\mathcal{C}=\{{\bf c}_1,{\bf c}_2,\ldots, {\bf c}_{16}\}$ in which the $\ell$-th codeword is defined as
\begin{equation*}
{\bf c}_\ell=\left[\phi_2\left(\hv_{1}^{\transp}\xv(g(\ell))\right),\ldots, \phi_2\left(\hv_{4}^{\transp}\xv(g(\ell))\right)\right]^{\transp}\in\{0,1\}^4.
\end{equation*} For example, when ${\bf H}={\bf I}_{4\times 4}$, then the minimum distance of the code $\mathcal{C}$ is one, i.e., $d_{\rm min}(\Hm ={\bf I}_{4\times 4})=1$. \\ 

\subsubsection{Effective channel} \label{des:effective_ch}
As shown in Fig.~\ref{e_model}, the effective channel consists of $N$ parallel $p$-ary input/output channels with input $\tv=[t_1,\ldots,t_{N}]^{\transp}$ and output $\rv=[r_1,\ldots,r_{N}]^{\transp}$. For the $i$-th subchannel, the transition probabilities, depending on users' messages $\wv=g(\ell)$,  are defined as
\begin{equation}
p_{\ell,i,j}\eqdef\PP(r_i= j |t_i=c_{\ell,i}),\label{eq:trans_prob}
\end{equation} for $j \in \{0,...,p-1\}$. From the $p$-level quantizer in (\ref{eq:p-ADC}) and the channel model in (\ref{eq:obs1}), we are able to compute the above transition probability using Q-function as
\begin{align}
p_{\ell,i,j}&=\PP\left( \Delta_{j} \leq \hv_i^{\transp}\xv(\wv) + z_i < \Delta_{j-1}\right),\nonumber\\
&=Q\left(2(\hv_i^{\transp}\xv(\wv)-\Delta_{j-1} )\right) +Q\left(2(\Delta_j  - \hv_i^{\transp}\xv(\wv)) \right),\label{eq:transit_prob}
\end{align} where 
\begin{equation*}
Q(t) = \frac{1}{2\pi}\int_{t}^{\infty} \exp\left(-\frac{t^2}{2}\right) dt.
\end{equation*} Using this, we define:

 \begin{definition}\label{def:trans_prob} For a given codeword index $\ell \in \Ic$ (i.e., users' message $\wv=g(\ell)$), we define a  {\em transition probability matrix} by $\Pm_{\ell}\in[0,1]^{N \times p}$ where the $(i,j)$-th element $\Pm_{\ell}(i,j)$ is 
 \begin{equation}
 \Pm_{\ell}(i,j)=p_{\ell,i,j}.
 \end{equation} 
 \hfill$\Diamond$
 \end{definition} 
The effective channel is fully characterized by the collection of transition probability matrices $\{\Pm_{\ell}: \ell \in\Ic\}$. We notice that the effective channel is in general not a symmetry channel. Even for the case of full CSI at the BS, the BS is not aware of the channel information of the effective channel as it does not know the user's messages.

\subsubsection{Decoding function} 
Using the above equivalent communication model, the multiuser massive MIMO detection problem is converted into the equivalent channel coding problem. Since the code $\Cc$ has been already constructed, our goal is to devise a decoding function to reliably estimate users' messages $\hat{\wv}$. 

We first review two well-known decoding methods: {\em minimum distance} decoding (MDD) and maximum likelihood (ML) decoding (MLD). MDD has been widely used in conventional channel coding problems \cite{MacWilliams}. When using MDD, user's messages $\hat{\wv}=g(\hat{\ell})$ are decoded by selecting the codeword that has the minimum Hamming distance from received signal vector ${\bf r}$, namely,
\begin{equation}
\hat{\ell}= \argmin_{\ell \in\Ic} d_{\rm h} (\rv, \cv_{\ell}).
\end{equation} Notice that MDD does not require the knowledge of the effective channel (i.e., transition probabilities).

MLD is an optimal decoding method under the premise that all transition probabilities $\{\Pm_{\ell}: \ell \in \Ic\}$ are perfectly known to the BS. When using MLD,  users' messages $\hat{\wv}=g(\hat{\ell})$ are decoded by choosing the codeword that maximizes the product of the transition probabilities, namely,
\begin{equation}
\hat{\ell}=\argmax_{\ell \in \Ic} \prod_{i=1}^{N} \Pm_{\ell}(i, r_i).\label{eq:ML}
\end{equation}  

For the two decoding methods, there exists the tradeoff between the detection error performance and the computational complexity; MLD can outperform MDD, while the latter requires much less channel state information than the former. They will be compared more rigorously in Section~\ref{sec:WMD}.

\vspace{0.1cm}
{\bf Remark 1:}  We want to notice that MLD was presented in \cite{Choi} using an original MIMO channel while the proposed MLD was developed for the effective channel defined in Section~\ref{des:effective_ch}. To distinguish these two methods, the proposed MLD is referred to as eMLD in the sequel. Namely, in this paper, eMLD implies the maximum-likelihood detection for the resulting $N$ parallel channels.
\vspace{0.1cm}

\subsection{Minimum Distance and Channel Goodness}\label{subsec:DG}

\begin{figure}
\centerline{\includegraphics[width=9cm]{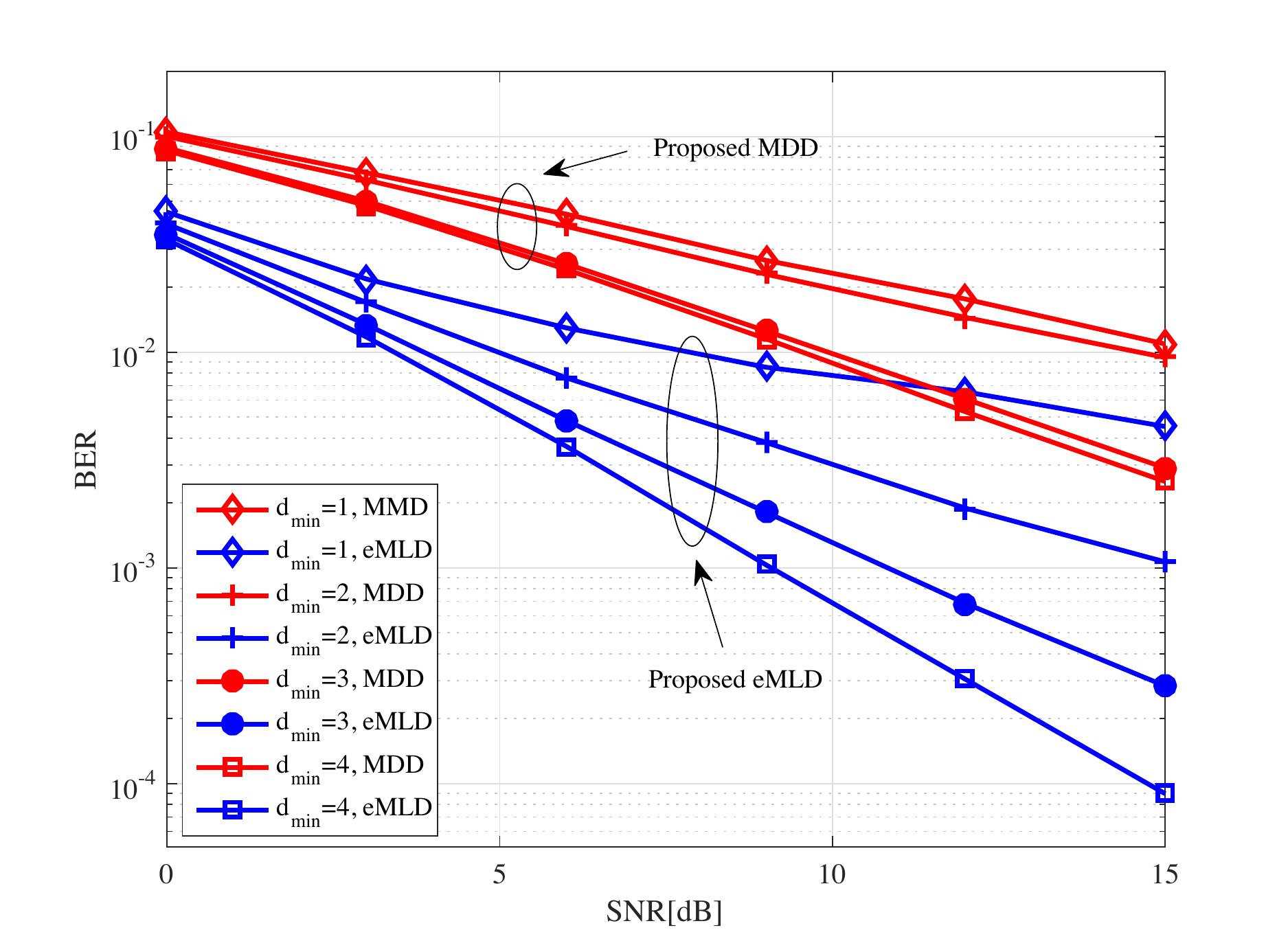}} \vspace{-0.5cm}
\caption{$K=2$, $N_{\rm r}=9$, and $p=2$. The BER performances as a function of a minimum distance of $\Cc$ (i.e., $d_{{\rm min}}(\Hm)$).}
\label{diversity_gain} 
\end{figure}

\begin{figure}
\centerline{\includegraphics[width=9cm]{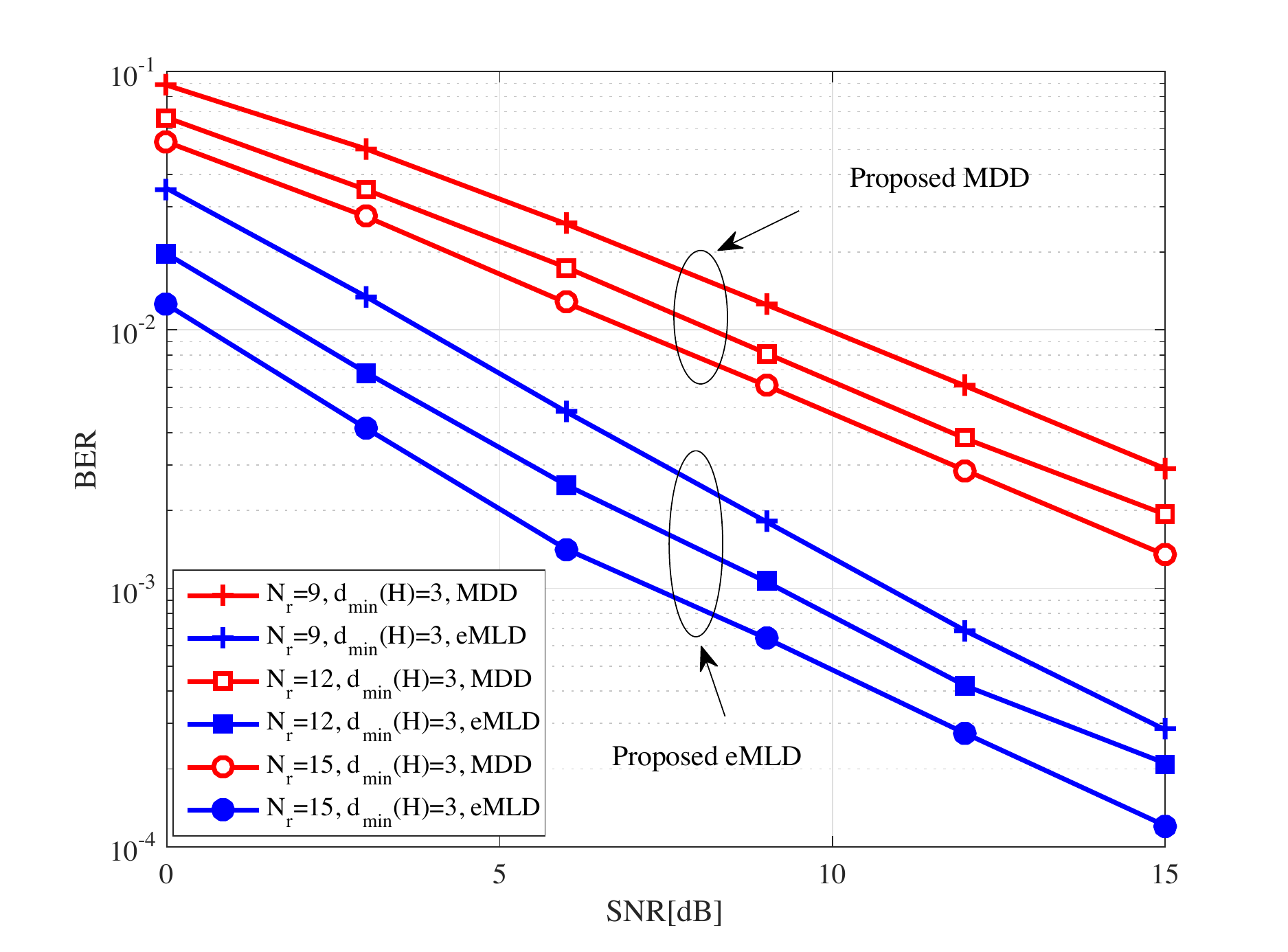}}\vspace{-0.5cm}
\caption{$K=2$ and $p=2$. Performance comparison of the proposed coding methods as a function of $N_{\rm r}$ when a random channel matrix yields the same minimum distance of the associated code $\Cc$.}
\label{power_gain}
\end{figure}

From the coding-theoretic viewpoint, we explain how to measure the goodness of a channel matrix for MIMO systems with low-resolution ADCs as a condition number for conventional MIMO systems. In coding theory, a minimum distance of a code plays a fundamental role in determining the goodness of the code since it determines the code performance especially at high SNRs \cite{MacWilliams}. Likewise, the performance of the proposed MIMO detection methods can be affected by the minimum distance of the channel-dependent code $\Cc$. We first show the impact of the minimum distance of the $\Cc$ on the error-performances. For the simplicity, we first focus on the MIMO systems with one-bit ADCs with
\begin{equation}\label{eq:one-bit}
\phi_2(u)\eqdef
\begin{cases}
0 & \mbox{ if } u \geq 0\\
1 & \mbox{ if } u  < 0.
\end{cases}
\end{equation} This is obtained from (\ref{eq:p-ADC}) with $\Delta_0 = 0$ (i.e., zero-threshold comparator). In this case, the effective channel is composed of $N$ parallel binary input/output channels with the transition probabilities of the $i$-th subchannel
\begin{equation}
p_{\ell,i,j} = 
\begin{cases}
\epsilon_{\ell,i} & \mbox{ if } j\neq i\\
 1 - \epsilon_{\ell,i} & \mbox{ if } j=i,
\end{cases}\label{eq:cross}
\end{equation}
where the crossover probability $\epsilon_{\ell,i}$, when users' messages $\wv=g(\ell)$ are sent, is computed from (\ref{eq:transit_prob}) as
\begin{equation}
\epsilon_{\ell,i} = Q\left(2|\hv_i^{\transp}\xv(g(\ell))|\right).
\end{equation} For the simulations, Rayleigh-fading channels are considered where each element of channel matrix ${\bf \tilde H} \in \CC^{N_{\rm r} \times K}$ is drawn from an independent and identically (IID) complex Gaussian random variables with zero-mean and unit-variance, and QPSK constellation is assumed. Let $\Hc$  denote the sample space containing all possible channel realizations $\Hm \in \RR^{N\times 2K}$. Then, we define:
\begin{equation}
\Hc_{d} \eqdef \{\Hm \in \Hc: d_{{\rm min}}(\Hm) = d\} \subseteq \Hc,
\end{equation} which contains all the channel realizations such that the corresponding codes have the minimum distance $d$, i.e., $d_{\rm min}(\Hm) = d$ for all $\Hm \in \Hc_{d}$.  To see the impact of the minimum distance on the code performance, we measure the conditional bit-error rate (BER), defined as
\begin{equation}
P_{\rm e}(d_{{\rm min}}(\Hm)=d)=\frac{1}{K}\sum_{k=1}^{K} P(\hat{w}_k \neq w_k | \Hc_d). \label{eq:cond_BER}
\end{equation}  From Fig.~\ref{diversity_gain}, we observe that the slope of BER, obtained by eMLD, is improved as the minimum distance of  $\Cc$ (i.e., $d_{{\rm min}}(\Hm)$) increases. Meanwhile, the slope of BER, obtained by MDD, seems to be related to the error-correction capability \cite{MacWilliams}, defined as
\begin{equation}
D=\left\lfloor \frac{d_{{\rm min}}(\Hm) - 1}{2} \right\rfloor.
\end{equation} Because of this difference, eMLD significantly outperforms the MDD and the performance gap of the two decoding methods becomes a larger as either $d_{\rm min}(\Hm)$ or $\SNR$ increases. This shows that the necessity of using proper weights in decoding to capture the different reliabilities of parallel subchannels. Also, it is expected that when the number of receiver antennas increases (equivalently, the corresponding minimum distance increases), the performance gain of eMLD would be remarkable. Not surprisingly, it is shown in Fig.~\ref{power_gain} that without enhancing the minimum distance, the only increase of the number of receiver antennas yields a $\SNR$ gain, i.e., it cannot improve the slope of BERs. From this analysis, we can see that the minimum distance of the $\Cc$ plays a crucial role in determining the performances of both decoding methods. Since the minimum distance of $\Cc$ is fully determined as the channel matrix $\Hm$, we can measure the {\em goodness} of a channel matrix using the minimum distance of its associated code $\Cc$. In other words, $d_{\rm min}(\Hm)$ can be used to evaluate the goodness of the $\Hm$ for MIMO systems with low-resolution ADCs as a condition number for conventional MIMO systems. Thus, we can say that a channel matrix $\Hm \in \Hc_{i}$ is a better channel than a channel matrix $\Hm' \in \Hc_{j}$ if $i>j$.\\

{\bf Remark 2:} The minimum distance $d_{\rm min}(\Hm)$ is not controlled and is determined as a function of a channel matrix $\Hm$. Hence, one might be interested in deriving the probability distribution of $d_{\rm min}(\Hm)$ (denoted by $f_{d_{\rm}(\Hm)}$) for any given probability distribution of $\Hm$ because it can enable us to estimate the performance of the MIMO systems. Recently, in 
\cite{Jeon_Hong_Lee}, $f_{d_{\rm}(\Hm)}$ is approximately derived for some simple case when BPSK constellation and Rayleigh-fading channel are considered. Further, it was shown that when $K=2$, $d_{\rm min}(\Hm)$ converges to $N$ asymptotically. Unfortunately, it is generally difficult to derive the $f_{d_{\rm}(\Hm)}$ since $d_{\rm min}(\Hm)$ is defined by a complex non-linear function of a random matrix $\Hm$. Instead, we are only able to compute the expectation of $d_{\rm min}(\Hm)$  using Monte-Carlo simulation, which can enable us to estimate the average BER performances.
\vspace{0.2cm}

\begin{figure}
\centerline{\includegraphics[width=9cm]{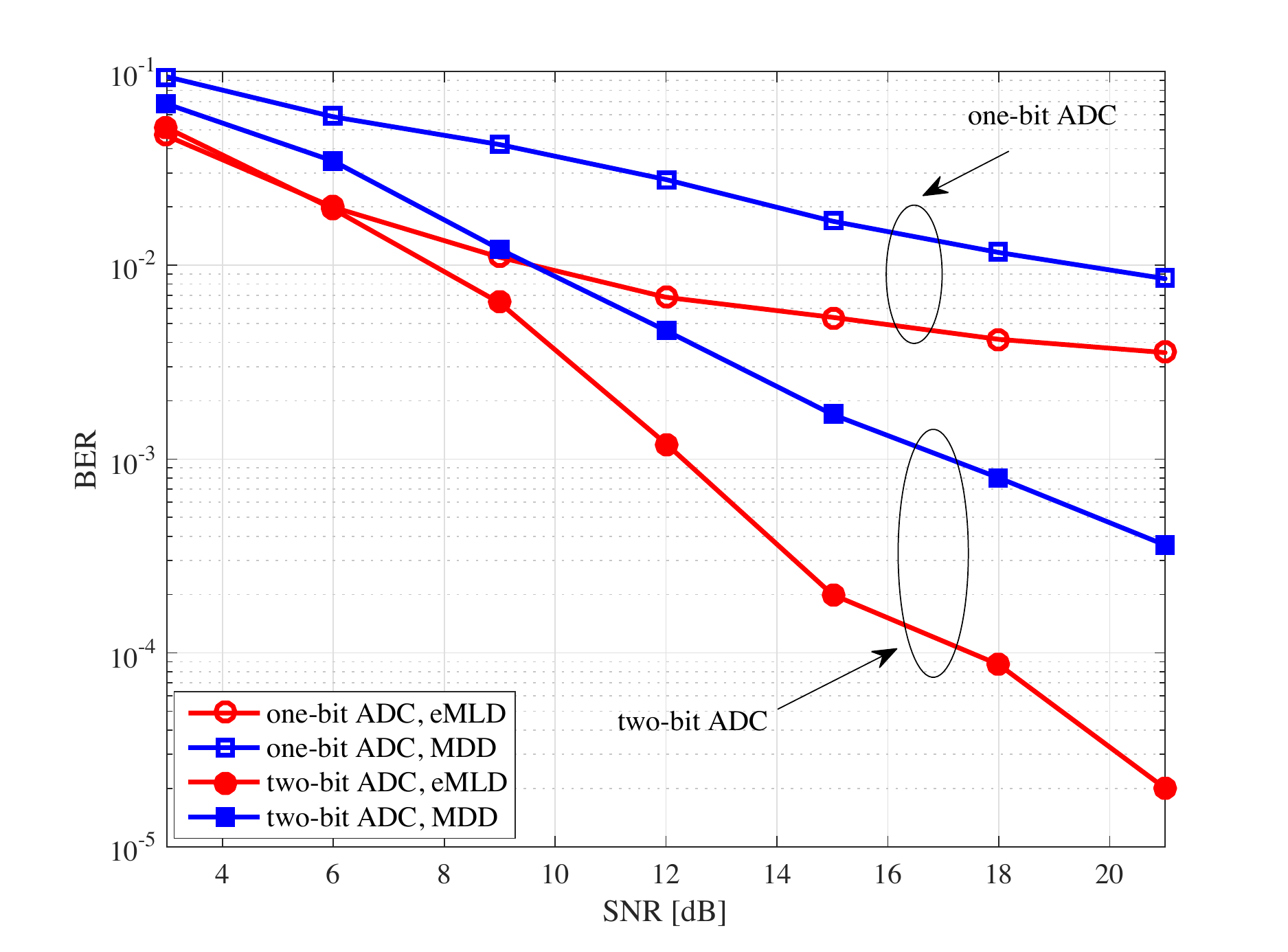}}\vspace{-0.5cm}
\caption{$K=2$, $N_{\rm r} = 6$. The BER performances of the proposed coding method as a function of ADC levels.}
\label{resolution}
\end{figure}

One can expect that the use of multi-bit ADCs improves the performance of the uplink MIMO systems. This is clearly explained from the coding-theoretic framework. Using the multi-bit ADCs, it can create a non-binary code $\Cc$ with a larger alphabet, which can typically have a larger minimum distance than a binary code. This will be verified by comparing the performances of the proposed methods for one-bit and two-bit ADCs. The quantization function of two-bit ADCs is defined as
\begin{equation}\label{eq:two-bit}
\phi_4(u)\eqdef
\begin{cases}
0 & \mbox{ if } u \geq \sqrt{\SNR}\\
1 & \mbox{ if } 0\leq u  < \sqrt{\SNR}\\
2 & \mbox{ if } - \sqrt{\SNR}\leq u <0\\
3, &\mbox{ if } u\leq - \sqrt{\SNR}.
\end{cases}
\end{equation} Consider the case of $K=2$ and $N_{\rm r}=6$ in which the corresponding blocklength of the code $\Cc$ is short as $N=12$. As explained in Remark 2, we can numerically compute the average minimum distance for both binary and quaternary code which correspond to 
 $\EE_{\Hm}[d_{\rm min}(\Hm)]=1.8$ and $\EE_{\Hm}[d_{\rm min}(\Hm)]=4.5$, respectively. Fig.~\ref{resolution} shows that this increment indeed improves the BER performance. Not surprisingly, the performance gap grows as SNR increases, namely, the increment of the minimum distance enhances the slope of BERs. As can be seen in Fig.~\ref{resolution}, the two-bit ADC in (\ref{eq:two-bit}) cannot achieve the better performance at lower SNRs; this failure implies that the minimum distance may not be an important factor at lower SNRs, as in conventional coding theory \cite{MacWilliams}. Hence, it would be a good research topic to develop a proper cost function to maximize the performance at target SNR; this is left for a future work.




\section{Weighted Minimum Distance Decoding}\label{sec:WMD}

In this section we first identify the fundamental difference between eMLD and MDD. Then, we will explain the rationale for that eMLD is significantly better than MDD in our problem unlike the conventional channel coding problems in which they are equivalent. Motivated by this, we present a novel {\em weighted} MD decoding (wMDD) as a practical approximate solution of eMLD.

\subsection{Comparison between eMLD and MDD}\label{subsec:MD_ML}

It is well-known that MDD is equivalent to eMLD in numerous channel coding problems \cite{MacWilliams}, in which all the information symbols of a codeword pass through the statistically identical channels with the same crossover probability.  Our problem, however, differs from the conventional channel coding problem in that each information symbol of a codeword passes through different channels, each with distinct crossover probabilities. This difference makes eMLD  to provide an unbounded gain over MDD in our problem as shown in Figs.~\ref{diversity_gain} and~\ref{power_gain}. To provide a more clear understanding of such difference, it is instructive to consider the simple case of  $p=2$ (i.e., one-bit ADCs). Assuming the users' message vector ${\bf w}_{\ell}=g(\ell)$, the modulated symbol vector $f(\wv) \in \mathcal{S}^K$  is transmitted. In this case, the transition probabilities of the $N$ subchannels are completely determined by the crossover probabilities $\epsilon_{\ell,i}$ for $i \in \{1,\ldots,N\}$. We first define an extended notion of Hamming distance:
\begin{definition}\label{def:w_MD} 
For any two vectors $\xv$ and $\yv$ of length $N$, we define a {\em weighted} Hamming distance $d_{wh}(\xv,\yv)$  with the weights $\{\alpha_i\}_{i=1}^{N}$ and $\{\beta_{i}\}_{i=1}^{N}$ as
\begin{equation*}
d_{\rm wh}(\xv,\yv;\{\alpha_i\},\{\beta_i\}) \eqdef \sum_{i=1}^{N} \alpha_i \mathbf{1}_{\{x_i = y_i\}} + \sum_{i=1}^{N} \beta_i \mathbf{1}_{\{x_i \neq y_i\}},
\end{equation*} where $\mathbf{1}_{\{\Ac\}}$ represents an indicator function with $\mathbf{1}_{\{\Ac\}} = 1$ if $\Ac$ is true, and  $\mathbf{1}_{\{\Ac\}} = 0$, otherwise. 
Notice that the Hamming distance is a special case of the weighted Hamming distance with the weights $\alpha_{i} = 0$ and $\beta_{i}=1$ for all $i \in \{1,...,N\}$.
\hfill$\Diamond$\\
\end{definition}

From Definition~\ref{def:w_MD},  both eMLD and MDD  can be represented in an unified view as
\begin{equation}
\hat{\ell} = \argmin_{\ell \in \Ic} d_{\rm wh} (\rv, \cv_{\ell};\{\alpha_{\ell,i}\},\{\beta_{\ell,i}\}),
\end{equation} with the different weight assignments. The weights when using eMLD are
\begin{align}
\alpha_{\ell,i} &= \log{(1-\epsilon_{\ell,i})^{-1}}\\
\beta_{\ell,i} &= \log{\epsilon_{\ell,i}^{-1}},
\end{align} for $i \in \{1,...,N\}$, which are obtained from (\ref{eq:ML}) and (\ref{eq:cross}). On the other hand, when using MDD, the weights are set to be $\alpha_{\ell,i} = 0$ and $\beta_{\ell,i} = 1$ for all $\ell \in \Ic$ and $i \in \{1,...,N\}$. This difference in the assignment of the weights clearly reveals that it is required to allocate a higher belief (or larger weights) for the received observation from more reliable subchannels. Namely, eMLD assigns proper soft weights according to the channel reliabilities. Whereas, MDD assigns hard weights, which does not contain the channel reliabilities adequately.\\

{\bf Example 2:} Suppose the single-user transmission scenario in which $K=1$, $N_{\rm r}=1$, $p=2$. When the user sends a QPSK symbol, it is assumed that there exists a channel and SNR, which can create an equivalent two subchannels, each with cross probabilities, $\epsilon_{\ell,1} = 10^{-1}$ and $\epsilon_{\ell,2} = 10^{-2}$. In this case, the  weights for eMLD are computed as
\begin{equation*}
\{\alpha_{\ell,1} = 0.15, \alpha_{\ell,2}=0.0145\},
\end{equation*}and
\begin{equation*}
\{\beta_{\ell,1}=3.32, \beta_{\ell,2}=6.64\}.
\end{equation*} It is  observed that a higher distance (i.e., weight) is allocated for a more reliable channel, for example, $\beta_{\ell,2} > \beta_{\ell,1}$ since $\epsilon_{\ell,2} < \epsilon_{\ell,1}$. As we explained, a higher belief is assigned to the information from more reliable channels. Because of such difference, eMLD  outperforms MDD as shown in Fig.~\ref{diversity_gain}. \\




\subsection{The Proposed  wMDD}\label{subsec:wMD}
Although MDD is suboptimal, it has a potential advantage with respect to implementation complexity. This is because MDD does not require to know the channel reliabilities of all subchannels. Motivated by this, we propose a new  {\em weighted} MDD (wMDD), which partially requires the unequal channel reliabilities of subchannels.  The proposed wMDD finds users' messages $\hat{\wv}=g(\hat{\ell})$ with
\begin{equation}
\hat{\ell} = \argmin_{\ell \in \Ic} d_{\rm wh}(\rv, \cv_{\ell};\{0\},\{\log{\epsilon_{\ell,i}^{-1}}\}),
\end{equation} where the weights are allocated by only the error probabilities of subchannels as
 
\begin{equation}
\epsilon_{\ell,i}= \sum_{j=0: j\neq \ell}^{p-1}\Pm_{\ell}(i,j).
\end{equation} Compared with eMLD, the proposed wMDD only needs to know error probabilities instead of all transition probabilities. This allows us to employ wMDD in a practical system where a BS has no knowledge of channel state information, since it is much simpler to estimate the error probabilities than all the transition probabilities accurately, with a limited training overhead  (see Section~\ref{sec:practical} for details). 

For the rest of this section, we will mathematically prove that wMDD outperforms MDD  essentially with the aid of proper weights. Before presenting a formal proof of our claim, we provide the basic idea of the proof using the simple case of $p=2$.\\

{\bf Example 3:} \label{ex:wMD} Suppose $\cv_{\ell}$ is a valid codeword. In this example, we focus on the error event that MDD (or wMDD) finds a wrong codeword  $\cv_{\ell'}$ for some $\ell'\neq \ell$.  Assuming $d_{h}(\cv_{\ell},\cv_{\ell'}) = 3$, we let $\{i_1,i_2,i_3\}$ denote the set of three positions that they differ.  Furthermore, let
\begin{equation}
\epsilon_{\ell,i_1} = 10^{-4}, \epsilon_{\ell,i_2} = 10^{-1}, \mbox{ and } \epsilon_{\ell,i_3} = 10^{-1}.\label{eq:cross_ex}
\end{equation} Notice that the $i_1$-th subchannel is more reliable than the other two subchannels. Since the error-probability only depend on the crossover probabilities of the above subchannels, we do not specify the others.

Let $\ev=[e_1,\ldots,e_N]^{\transp}$ represent an error vector, where $e_{i}=1$ indicates that the error occurs at the $i$-th subchannel. To compute the error-probability, we focus on the errors corresponding to the positions $\{i_1,i_2,i_3\}$. When MDD is used, an error-vector $\ev$ yields a wrong decision if $d_h([e_{i_1},e_{i_2},e_{i_3}], [0,0,0]) > 1$. From this, the collection of such error vectors under MDD is obtained as
\begin{align*}
&\Ac_{\rm MD}(\ell'|\ell)=\left\{\ev \in \{0,1\}^N: [e_{i_1},e_{i_2},e_{i_3}] \in \Rc_{\rm MD} \right\},
\end{align*} where $\Rc_{\rm MD} = \{[1,1,0],[1,0,1],[0,1,1],[1,1,1]\}$. Namely, if an error-vector $\ev \in \Ac_{\rm MD}(\ell'|\ell)$ occurs, MDD cannot find the valid codeword. Using the crossover probabilities in  (\ref{eq:cross_ex}), we can compute the error-probability as
\begin{equation*}
\PP(\Ac_{\rm MD}(\ell'|\ell))= 10^{-2}.
\end{equation*} Next, we consider the proposed wMDD. Using the crossover probabilities, we first compute the weighted Hamming distance between $\cv_{\ell}$ and $\cv_{\ell'}$ using the weights $\{\log{\epsilon_{\ell,i_j}^{-1}}\}$, which is given by $d_{\rm wmin} = 9.9658$. When wMDD is used, an error-vector $\ev$ yields a wrong decision if 
\begin{equation}
 \sum_{j=1}^{3} \left(\log{\epsilon_{\ell,i_j}^{-1}}\right)\mathbf{1}_{\{e_{i_j} \neq 0\}} > \frac{d_{\rm wmin}}{2}.
\end{equation} From this, the collection of  such error vectors under wMDD is obtained as
\begin{equation*}
\Ac_{\rm wMD}(\ell'|\ell)=\left\{\ev \in \{0,1\}^N: [e_{i_1},e_{i_2},e_{i_3}] \in\Rc_{\rm wMD}\right\},
\end{equation*}  where $\Rc_{\rm wMD}=\{[1,0,0],[1,1,0],[1,0,1],[1,1,1]\}$. The corresponding error-probability is computed as
\begin{equation*}
\PP(\Ac_{\rm wMD}(\ell'|\ell)) = 10^{-4}.
\end{equation*} In this example, we observe that some error vectors are not decodable by MDD but decodable by wMDD, and vice versa (i.e., $R_{\rm MD} \neq R_{\rm wMD}$), which makes it complicated to prove Theorem~\ref{thm:wMD}. The main reason why wMDD performs better than MDD is that the former error vectors occur with much lower probability than the latter error vectors, i.e., 
\begin{align}
&\PP\left(\Ac_{\rm wMD}(\ell'|\ell) \setminus \Ac_{\rm MD}(\ell'|\ell) \right) \nonumber \\
&\;\;\;\;\;\;\;\;\;\;\;\;\;\;\;\;\;\;\;\;\;\;\;\;\;\;\;  \ll\PP\left(\Ac_{\rm MD}(\ell'|\ell)\setminus \Ac_{\rm wMD}(\ell'|\ell)\right),\label{eq:comp}
\end{align} where
\begin{align}
&\Ac_{\rm wMD}(\ell'|\ell) \setminus \Ac_{\rm MD}(\ell'|\ell)\nonumber\\
&\;\;\;\;\;\;\;\;\;\;\;\;\;\;\;\;=\left\{\ev \in \{0,1\}^N: [e_{i_1},e_{i_2},e_{i_3}]=[1,0,0]\right\}\label{eq:set1}\\
&\Ac_{\rm MD}(\ell'|\ell)\setminus \Ac_{\rm wMD}(\ell'|\ell)\nonumber\\
&\;\;\;\;\;\;\;\;\;\;\;\;\;\;\;\;=\left\{\ev \in \{0,1\}^N: [e_{i_1},e_{i_2},e_{i_3}]=[0,1,1]\right\}.\label{eq:set2}
\end{align} This difference will be used as the underlying idea  for the proof of Theorem~\ref{thm:wMD}.\\

We first provide the following definition which will be used for the proof of Theorem~\ref{thm:wMD}.

\begin{definition} Consider a $p$-ary vector $\xv=[x_1,\ldots,x_N]^{\transp}$ with $x_i \in \{0,...,p-1\}$. Its complement, denoted by $\bar{\xv}$, is a length-$N$ vector obtained by replacing non-zero values in $\xv$ by zeros in $\bar{\xv}$, and zeros in $\xv$ by some non-zero values in $\bar{\xv}$. In general, a complement of $\xv$ is not unique and in fact, there are $(p-1)^{N-d_h(\xv)}$ number of different complements of $\xv$.\hfill$\Diamond$
\end{definition} 

The main theorem of this section is provided as follows.

\begin{theorem}\label{thm:wMD} Let $P_{\rm e, wMD}$ and $P_{\rm e, MD}$ represent the error-probability of wMDD and MDD, respectively. Then, we have:
\begin{equation}
P_{\rm e, wMD} \leq P_{\rm e, MD}.
\end{equation}
\end{theorem}


\begin{IEEEproof}  Recall that $\Ic$ denotes the index set of the codewords of $\Cc$. We let $\Ac_{\rm wMD}(\ell'|\ell)$ (resp. $\Ac_{\rm MD}(\ell'|\ell)$) denote the error-event that wMDD (resp. MDD) finds a wrong codeword $\cv_{\ell'}$ when  $\cv_{\ell}$ is a valid codeword (see (\ref{eq:error_event1}) (resp. (\ref{eq:error_event2})) for details). Then, the error probabilities of wMDD and MDD are respectively defined as
\begin{align}
P_{\rm e,wMD} &= \frac{1}{|\Ic|} \sum_{\ell \in\Ic}\sum_{\ell' \in \Ic:\ell'\neq \ell} \PP(\Ac_{\rm wMD}(\ell'|\ell))\\
P_{\rm e,MD} &= \frac{1}{|\Ic|} \sum_{\ell \in \Ic}\sum_{\ell'\in\Ic: \ell' \neq \ell} \PP(\Ac_{\rm MD}(\ell'|\ell)).
\end{align} 
For the proof, we will show that, for any index pair $(\ell,\ell'\neq \ell)$, 
\begin{equation}
\PP\left(\Ac_{\rm wMD}(\ell'|\ell)\right) \leq \PP\left(\Ac_{\rm MD}(\ell'|\ell)\right).\label{eq:obj_pf}
\end{equation} This immediately shows that $P_{\rm e, wMD} \leq P_{\rm e, MD}$. 

Now we focus on the proof of (\ref{eq:obj_pf}). Without loss of generality, we assume that $\cv_{\ell}$ and $\cv_{\ell'}$ are valid and wrong codewords, respectively. We let $\epsilon_{\ell,i}$, $i \in \{1,...,N\}$ denote the error-probability of each subchannel $i$, when the valid codeword $\cv_{\ell}$ is transmitted. Let $\{i_1,...,i_d\}$ denote the set of $d=d_{h}(\cv_{\ell},\cv_{\ell'})$ positions that they differ. The rest of this proof almost follows the procedures in Example 3. 

First of all, if there exists another codeword $\cv_{\ell''}$ for which the set of different positions from the $\cv_{\ell}$ is a proper subset of $\{i_1,...,i_d\}$, it is obvious that $\PP\left(\Ac_{\rm wMD}(\ell'|\ell)\right) = \PP\left(\Ac_{\rm MD}(\ell'|\ell)\right)=0$, since $\cv_{\ell'}$ should not be chosen with probability 1. This is the trivial case to show that (\ref{eq:obj_pf}) is satisfied.  

Next, we consider the case that there is no such codeword. Define the set of error vectors for which MDD finds the wrong codeword $\cv_{\ell'}$ as
\begin{align}
&\Ac_{\rm MD}(\ell'|\ell)\nonumber\\
& =\left\{\ev \in \{0:p-1\}^{N}: d_h([e_{i_1},...,e_{i_d}]) > \left\lfloor \frac{d-1}{2} \right\rfloor \right\}\label{eq:error_event1}.
\end{align} For wMDD, we  compute the weighted Hamming distance as
\begin{equation}
d_{\rm wmin} = \sum_{j=1}^{d} \log{\epsilon_{\ell,i_j}^{-1}}.
\end{equation} From this, we define the set of error vectors for which wMDD finds the wrong codeword $\cv_{\ell'}$ as
\begin{align}
&\Ac_{\rm wMD}(\ell'|\ell) \nonumber \\
&=\underbrace{\left\{\ev \in \Ac_{\rm MD}: \sum_{j=1}^{d} \left( \log{\epsilon_{\ell,i_j}^{-1}} \right) \mathbf{1}_{\{e_{i_j} \neq 0\}} > \frac{d_{\rm wmin}}{2} \right\}}_{\eqdef \Ac_{\rm wMD,1}} \nonumber \\
&\bigcup \underbrace{\left\{\ev \notin \Ac_{\rm MD}: \sum_{j=1}^{d} \left(\log{\epsilon_{\ell,i_j}^{-1}}\right) \mathbf{1}_{\{e_{i_j} \neq 0\}} > \frac{d_{\rm wmin}}{2} \right\}}_{\eqdef \Ac_{\rm wMD,2}}.\label{eq:error_event2}
\end{align} Clearly, we have $\Ac_{\rm wMD,1} \cap \Ac_{\rm wMD,2} = \phi$. Then, we can rewrite the error probabilities of wMDD and MDD as
\begin{align}
\PP(\Ac_{\rm wMD}(\ell'|\ell)) &=\PP(\Ac_{\rm wMD,1}) + \PP(\Ac_{\rm wMD,2})\label{eq:pf1}\\
\PP(\Ac_{\rm MD}(\ell'|\ell)) &=\PP(\Ac_{\rm wMD,1}) + \PP(\Ac_{\rm MD,2}), \label{eq:pf2}
\end{align} where for the simplicity of notation, we let $\Ac_{\rm MD,2}=\Ac_{\rm MD}(\ell'|\ell)\setminus \Ac_{\rm wMD,1}$. From (\ref{eq:pf1}) and (\ref{eq:pf2}), the proof is completed by showing that
\begin{equation}
\PP(\Ac_{\rm wMD,2}) \leq \PP(\Ac_{\rm MD,2}).\label{eq:r_pf}
\end{equation} 
Define the subset of  $\Ac_{\rm wMD,2}$ as
\begin{equation}
\Ac_{\rm wMD,2}(i) =\left\{\ev \in \Ac_{\rm wMD,2}: d_{\rm h}([e_{i_1},...,e_{i_d}]) = i \right\},
\end{equation} for $i =0,1,...,\left\lfloor\frac{d-1}{2} \right\rfloor $. They partition the $\Ac_{\rm wMD,2}$ since $\Ac_{\rm wMD,2}(i)$'s are mutually disjoint and 
\begin{equation}
\Ac_{\rm wMD,2}=\bigcup_{i=0}^{\left\lfloor\frac{d-1}{2} \right\rfloor}\Ac_{\rm wMD,2}(i).
\end{equation} We only focus on the positions $\{i_1,...,i_d\}$ by assuming that the other positions are fixed by some arbitrary values from $\{0,...,p-1\}$. For any fixed $i \leq \left\lfloor\frac{d-1}{2} \right\rfloor $ non-zero positions from $\{i_1,...,i_d\}$, there are the $(p-1)^{i}$ number of error vectors, denoted by $\ev_{1},...,\ev_{(p-1)^{i}} \in \Ac_{\rm wMD,2}(i)$. For such error vectors, there are the $(p-1)^{d-i} \geq (p-1)^{i}$ number of their complements. From Lemma~\ref{lem:thm} below, they all belong to $\Ac_{\rm MD,2}$. In addition, from (\ref{eq:lem}) and (\ref{eq:lem1}), we can see that
\begin{equation}
\PP(\{\ev\}) \leq \PP(\{\bar{\ev}\}),
\end{equation} for any $\ev \in \Ac_{\rm wMD,2}(i)$ and any its complement $\bar{\ev} \in \Ac_{\rm MD,2}$. Using this, we can obtain 
\begin{align*}
\PP(\{\ev_{1},...,\ev_{(p-1)^{i}}\})&=\sum_{j=1}^{(p-1)^{i}}\PP(\{\ev_{j}\})\\
& \leq \PP(\{\mbox{all their complements}\}).
\end{align*} Clearly, the above upper bound is satisfied for any other $i$ non-zero positions. In addition, for any two different $i$ non-zero positions, the corresponding their complement sets are disjoint. Therefore, we have:
\begin{equation}
\PP(\Ac_{\rm wMD,2}(i)) \leq \PP(\Bc_{i}),\label{eq:bound1}
\end{equation} where $\Bc_{i}$ denotes the collection of all the complements of error vectors with $i$ non-zero positions from $\{i_1,...,i_d\}$. Since we have that
\begin{equation}
\Bc_{i} \subset \Ac_{\rm MD,2}, \label{eq:bound2}
\end{equation} for $i=0,...\left\lfloor\frac{d-1}{2} \right\rfloor$,  and $\Bc_i$'s are disjoint, we obtain
\begin{align}
\PP(\Ac_{\rm wMD,2}) &= \sum_{i=0}^{\left\lfloor\frac{d-1}{2} \right\rfloor }\PP(\Ac_{\rm wMD,2} (i))\\
&\stackrel{(a)}{\leq}  \sum_{i=0}^{\left\lfloor\frac{d-1}{2}\right\rfloor} \PP(\Bc_{i}) \\
&\stackrel{(b)}{\leq} \PP(\Ac_{\rm MD,2}),
\end{align} where (a) and (b) are respectively from (\ref{eq:bound1}) and (\ref{eq:bound2}). This completes the proof of Theorem~\ref{thm:wMD}.
\end{IEEEproof}

\begin{lemma}\label{lem:thm} For any vector $\ev \in \Ac_{\rm wMD,2}$,  all the complements of the $\ev$ with respect to the positions of $\{i_1,\ldots,i_d\}$ belong to  $\Ac_{\rm MD}(\ell'|\ell)\setminus \Ac_{\rm wMD,1}$.
\end{lemma}
\begin{IEEEproof}  Let $\bar{\ev}$ be an arbitrary complement of $\ev$. By definition, we have
\begin{align}
&\underbrace{\sum_{j=1}^{d} \left( \log{\epsilon_{\ell,i_j}^{-1}}\right)  \mathbf{1}_{\{e_{i_j} \neq 0\}}}_{(a)}  + \underbrace{\sum_{j=1}^{d} \left(\log{\epsilon_{\ell,i_j}^{-1}}\right)  \mathbf{1}_{\{\bar{e}_{i_j} \neq 0\}}}_{(b)} =d_{\rm wmin},\label{eq:lem}
\end{align} since if $e_{i_j} = 0$, then $\bar{e}_{i_j} \neq 0$, and vice versa. Since $(a) > d_{\rm wmin}/2$, we obtain that 
\begin{equation}
(b) \leq \frac{d_{\rm wmin}}{2}.\label{eq:lem1}
\end{equation} Also, since $\ev \notin \Ac_{\rm MD}$, the $\bar{\ev}$ should satisfy the
\begin{equation}
\sum_{j=1}^{d} \mbox{1}_{\{\bar{e}_{i_j} \neq 0\}} > \left\lfloor \frac{d-1}{2} \right\rfloor.\label{eq:lem2}
\end{equation} From (\ref{eq:lem1}) and (\ref{eq:lem2}), we can conclude that $\bar{\ev} \in \Ac_{\rm MD}(\ell'|\ell)\setminus \Ac_{\rm wMD,1}$. This completes the proof.
\end{IEEEproof}

\subsection{Low-Complexity Approach} \label{subsec:discussion}

The major drawback of the proposed detection methods is that their decoding complexities grow exponentially with the number of uplink users. This fact prevents from the use of the proposed method in massive MIMO systems especially when the number of users in the network is large. The exactly same problem has been occurred in MLD \cite{Choi} and supervised learning approach (SRA) \cite{Jeon_Hong_Lee}. To resolve this problem, the complexity reduction techniques have been presented.  For example, in \cite{Choi}, the MLD problem was approximated as a convex optimization problem, which can be solved efficiently, and then using that solution, a simple symbol-by-symbol detection was performed. Recently in \cite{Jeon_Hong_Lee}, a successive interference cancellation (SIC) based method has been presented, in which input vectors corresponding to users' messages are partitioned into two subvectors and then they are detected successively in the manner of SIC.  In this approach, the decoding method is applied to the two MIMO systems, each with smaller number of users than $K$ and hence the overall decoding complexity is reduced. Furthermore, it was shown that by carefully partitioning the input vectors, this approach can almost achieve the performance of the original detection method.

The SIC idea in  \cite{Jeon_Hong_Lee}  can be immediately applied to the proposed wMDD (also, eMLD and MDD). Specifically, the input vector of length $K$ is partitioned into two subvectors whose lengths are respectively $K_1$ and $K_2$ with $K_1+K_2=K$.  Here, the partition is performed so that the distance of the subspaces, which are spanned by the submatrices of $\Hm$ corresponding to the two subvectors (denoted by $\Hm_1$ and $\Hm_2$), is maximized (see \cite{Jeon_Hong_Lee} for the detailed algorithm). Then, we are able to construct code $\Cc_{i}$ using submatrix $\Hm_i$.  With these codes, the decoding is performed in a successive manner as follows:
\begin{enumerate}
\item Using the code $\Cc_1$ and the received observations, wMDD finds some part of users' messages corresponding to the subvector $i$.
\item The effect of the estimated subvector 1 is eliminated from the observations (see  \cite{Jeon_Hong_Lee} for details). Then, using the code $\Cc_2$ and the improved observations, wMDD finds the remaining part of users' messages. 
\end{enumerate} Using the above technique, the decoding complexity is reduced from $m^K$ to $m^{K_1}+m^{K_2}$. For example, when $m=4$ (e.g., QPSK modulation) and $K=12$,  the original wMDD requires the  $4^{12}$ distance comparisons. Whereas, using the reduction technique with  $K_1 = 6$ and $K_2 = 6$, the wMDD requires the $2\times 4^6$ distance comparisons.  In general, the decoding complexity and the detection performance depend on the choices of $K_1$ and $K_2$ subject to $K=K_1+K_2$. Since there are a lot of possible choices especially for a large $K$, we need to optimize them so that the performance is maximized subject to an affordable decoding complexity.

\section{Practical Communication Method \\Using the Proposed wMDD} \label{sec:practical}


\begin{figure}
\centerline{\includegraphics[width=9cm]{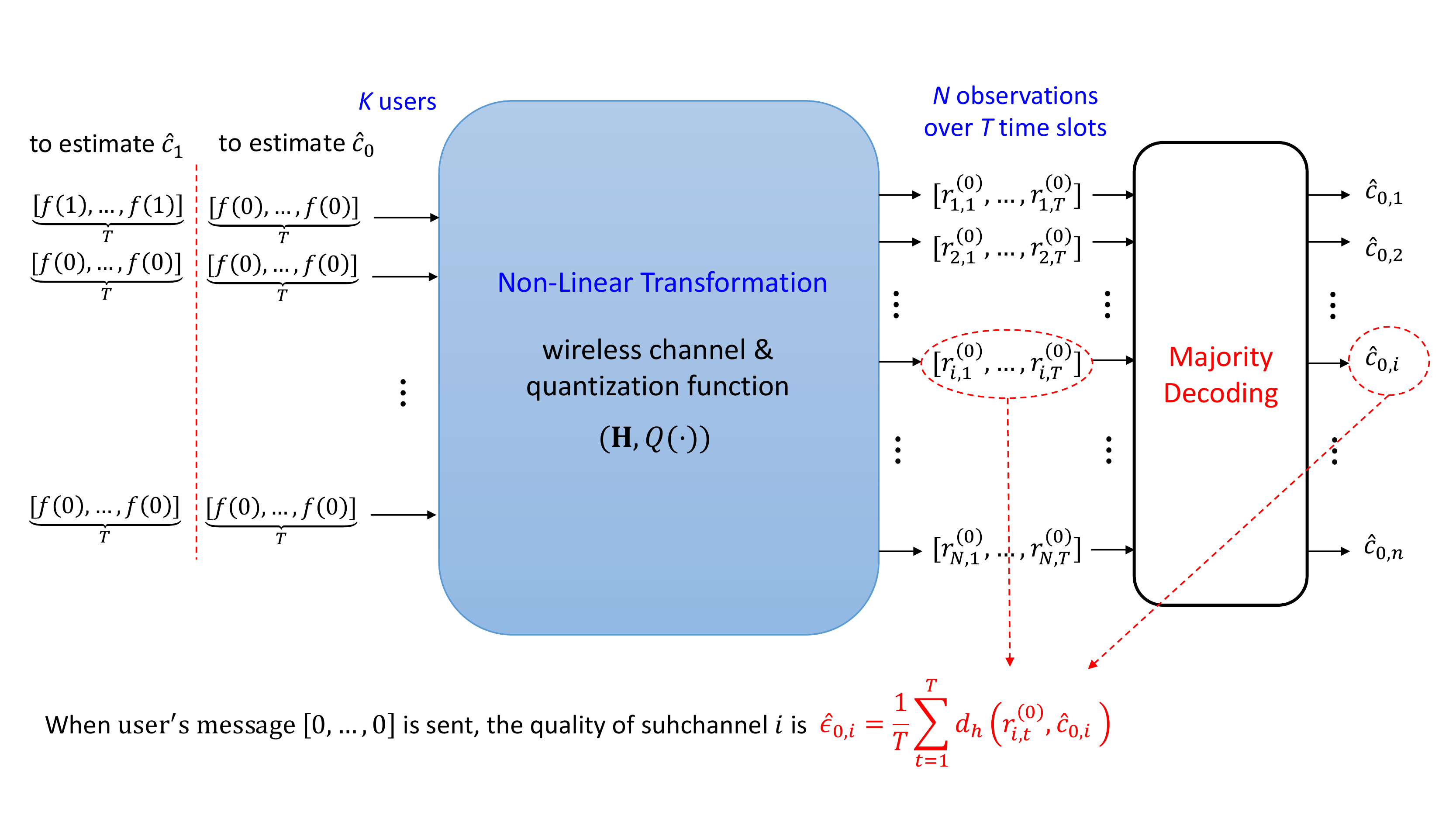}}
\caption{Illustration of an implicit channel training method.}
\label{t_model}
\end{figure}

Assuming the perfect CSIR at the BS, in Section~\ref{sec:Main} and~\ref{sec:WMD}, we developed the novel multiuser MIMO detection methods inspired by coding theory. In practical systems, however, it is not reasonable to assume to perfect CSIR because CSI is typically estimated at the BS from uplink pilot transmissions. In this section, we propose the channel training methods suitable for the proposed wMDD in which a code  $\Cc$ (i.e., $m^K$ codewords of $\Cc$) and an error-probability of each subchannel are estimated. We assume a block-fading channel where the channel is static during a channel coherence interval, i.e., $T_{\rm c}$ time slots in a given fading block and changes the channel independently from block-to-block. For such channel, we propose two channel training methods which are respectively called  {\em implicit} and {\em explicit} channel training methods. Their major difference is that the explicit method only needs an explicit channel estimation process (e.g., estimating a channel matrix $\Hm$). The detailed channel training procedures of the both methods are described below.

 \subsection{Implicit Channel Training Method}\label{subsec:Implicit}
 
 In this method, the uplink users repeatedly send all possible input vectors so that BS observes the multiple received signals. These multiple observations enable the BS to estimate the codewords of $\Cc$ and the reliabilities of subchannels (see Fig.~\ref{t_model}). The advantage of this method is the absence of channel estimation process during the channel training. This advantage is particularly attractive for a MIMO system with low-level ADCs (e.g., one-bit ADCs) because it is difficult to obtain an accurate CSIR using conventional pilot-based channel estimation techniques.

Let the first $T_{\rm o}=|\Cc|T< T_{\rm c}$ time slots be devoted for a training phase and the remaining $T_{\rm d}= T_{\rm c}-|\Cc|T$ time slots be dedicated to a data transmission phase. During the training phase,  all the codewords $\cv_{0},\cv_{1},\ldots,\cv_{m^K -1}$ are estimated in that order. Let  $\ell=b_0m^0+b_1m^1+\cdots+b_{K-1}m^{K-1}$. For each codeword $\cv_\ell \in \Cc$, each user $k$ transmits a $m$-ary training sequence as
\begin{equation}
\underbrace{[b_{k-1},b_{k-1},...,b_{k-1}]}_{T} 
\end{equation} during $T$ time slots. Since we use the $T$ time slots to estimate each codeword, the overall training overhead is
\begin{equation}
T_{\rm o} = T\times m^K. \label{eq:t_overhead}
\end{equation} From the $i$-th receiver antenna, the BS observes the 
\begin{equation}
\rv_{i}^{(\ell)}=[r_{i,kT+1},\ldots,r_{i,(k+1)T}]^{\transp}, \label{eq:trained}
\end{equation} for $i \in \{1,...,N\}$. From this, the $i$-th element of  $\cv_\ell$ (denoted by $c_{\ell,i}$) is estimated using the simple {\em majority rule} as
\begin{equation}
\hat{c}_{\ell,i} = \argmax_{b \in \{0:p-1\}} \Nc_{b}(\rv_{i}^{(\ell)}), \label{eq:est_codeword}
\end{equation} where $\Nc_{b}(\rv)$ represents the number of $b$'s in $\rv$. Repeating the above procedures for $\ell=0,1,\ldots, m^K-1$, we are able to estimate a $p$-ary code $\hat{\Cc}$ as
\begin{equation}
\hat{\Cc}=\{\hat{\cv}_0,\ldots,\hat{\cv}_{m^K-1}\},\label{eq:est_code}
\end{equation} where $\hat{\cv}_\ell = [\hat{c}_{\ell,1},\ldots,\hat{c}_{\ell,N}]^{\transp}$. The proposed MDD can be performed using the estimated code $\hat{\Cc}$.

Next, for eMLD, we must estimate  all transition probabilities $\{\Pm_{\ell}: \ell \in \Ic\}$ from the above training observations. To obtain an accurate  $\{\Pm_{\ell}: \ell \in \Ic\}$, the training overhead should be very large. In contrast, for wMDD, we only need to estimate the error-probability of each subchannel. This can be simply performed using the estimated codewords and the training observations as
\begin{equation}
\hat{\epsilon}_{\ell,i} = \frac{1}{T}\sum_{t=1}^{T}d_{\rm h}(\hat{c}_{\ell,i},r_{i,(kT+t)}).\label{eq:est_perror}
\end{equation} From (\ref{eq:est_code}) and (\ref{eq:est_perror}), the proposed wMDD finds users' messages $\hat{\wv} = g(\hat{\ell})$  as
\begin{equation}\label{eq:w_MD}
\hat{\ell}= \argmin_{\ell \in \Ic} d_{\rm wh}(\rv, \hat{\cv}_{\ell}:\{0\},\{\log{\hat{\epsilon}_{\ell,i}^{-1}}\}).
\end{equation}

\vspace{0.2cm}
{\bf Remark 3:} When a training overhead is small (e.g., $T$ is small), an empirical error-probability in (\ref{eq:est_perror}) can be estimated as zero, although the corresponding subchannel should not be a perfect channel. This overestimation can result in a severe error-floor problem. To overcome this numerical problem, we assign a minimum value to the empirical error-probability for the purpose of implementation. For the simulations, we selected the minimum value as $10^{-3}$.

\subsection{Explicit Channel Training Method}\label{subsec:Explicit}

The major limitation of the implicit channel training method is that the training overhead in (\ref{eq:t_overhead}) grows exponentially with the number of users $K$. Thus, the implicit method is not suitable for the systems consisting of lots of active users. To overcome this, we propose an explicit channel training method in which the BS alone artificially generate the received signals for all possible input vectors. This is performed using an estimated channel matrix $\hat{\Hm}$. Then the generated observations can take the role of the training observations in (\ref{eq:trained}) of the implicit channel training method. Since this method is enable only when the BS knows the channel matrix, the explicit channel training method requires a conventional channel estimation process (see \cite{Choi,Li} for more details), which is the main difference from the implicit channel training method. The explicit channel training method is composed of the two parts as channel estimation process and artificial training as 
\begin{enumerate}
\item The BS first estimates a channel matrix $\hat{\Hm}$ using the conventional channel estimation method with $T_{\rm t}$ pilot signals (see \cite{Choi,Li} for the detail channel estimation methods)
\item Using the estimated channel matrix $\hat{\Hm}$, the BS alone follows the procedures in Fig.~\ref{t_model}. Namely, it  generates the artificial observations for each input vector as
\begin{equation}
\hat{\rv}_{i}^{(\ell)}=[\hat{r}_{i,kT+1},\ldots,\hat{r}_{i,(k+1)T}]^{\transp},
\end{equation} for $i \in \{1,...,N\}$. Then, it estimates the codewords and the error probabilities using (\ref{eq:est_codeword}) and (\ref{eq:est_perror}), respectively.
\end{enumerate} In this method, the overall training overhead is $T_{\rm o} = T_{\rm t}$, which is not necessarily scaled with the number of users as in the implicit channel training method.


\subsection{Numerical Results}\label{sec:numerical}

In this section, we evaluate the performances of the proposed detection methods in Section~\ref{subsec:Implicit} and~\ref{subsec:Explicit}, compared to the existing MIMO detection techniques. Rayleigh-fading channels are assumed where each element of the channel matrix $\Hm$ is drawn from an IID circularly-symmetric complex Gaussian random variable with zero mean and unit variance. A block fading duration (i.e., a coherence time interval) is set to be  $T_{\rm c} = T_{\rm o} + T_{\rm d}=1000$ and the training overhead is limited by the $10\%$ of the coherence time (i.e., $T_{\rm o} \leq 100$). In addition, the same transmit power is assigned to the both training and data transmissions, i.e., there is no power-boosting for pilot sequences. For the implicit channel training method, we further reduce the training overhead using the reduction technique in \cite{Jeon_Hong_Lee}. The main idea of this technique is to exploit the symmetry of the transmit constellation and the quantization function in ADCs. To be specific, during the $T\times m^K/2$ time slots, the BS first observes the $\{\rv_{i}^{(0)},...,\rv_{i}^{(\frac{m^K}{2}-1)}: i=1,...,N\}$. Then, without actual transmissions from the users, the BS produces the 
\begin{equation}
\rv_{i}^{(m^K - \ell-1)} = - \rv_{i}^{(\ell)}
\end{equation} for $\ell=0,1,...,\frac{m^K}{2}-1$ and $i=1,...,N$. Thus, the training overhead of the implicit method is reduced by the half as
\begin{equation}
T_{\rm o} = \frac{T \times m^K}{2}. 
\end{equation} For the explicit channel training method, ZF-type CE in \cite{Choi} is used to estimate a channel matrix and $T=25$ is used for artificial channel training.

For the comparisons, we consider the existing detection methods as MLD and ZFD in \cite{Choi}, and SLA in \cite{Jeon_Hong_Lee}. In this section, we compare the performances of all the detection techniques with their best performances, without applying the complexity-reduction techniques in Section~\ref{subsec:discussion}.

\begin{figure}
\centerline{\includegraphics[width=9.5cm]{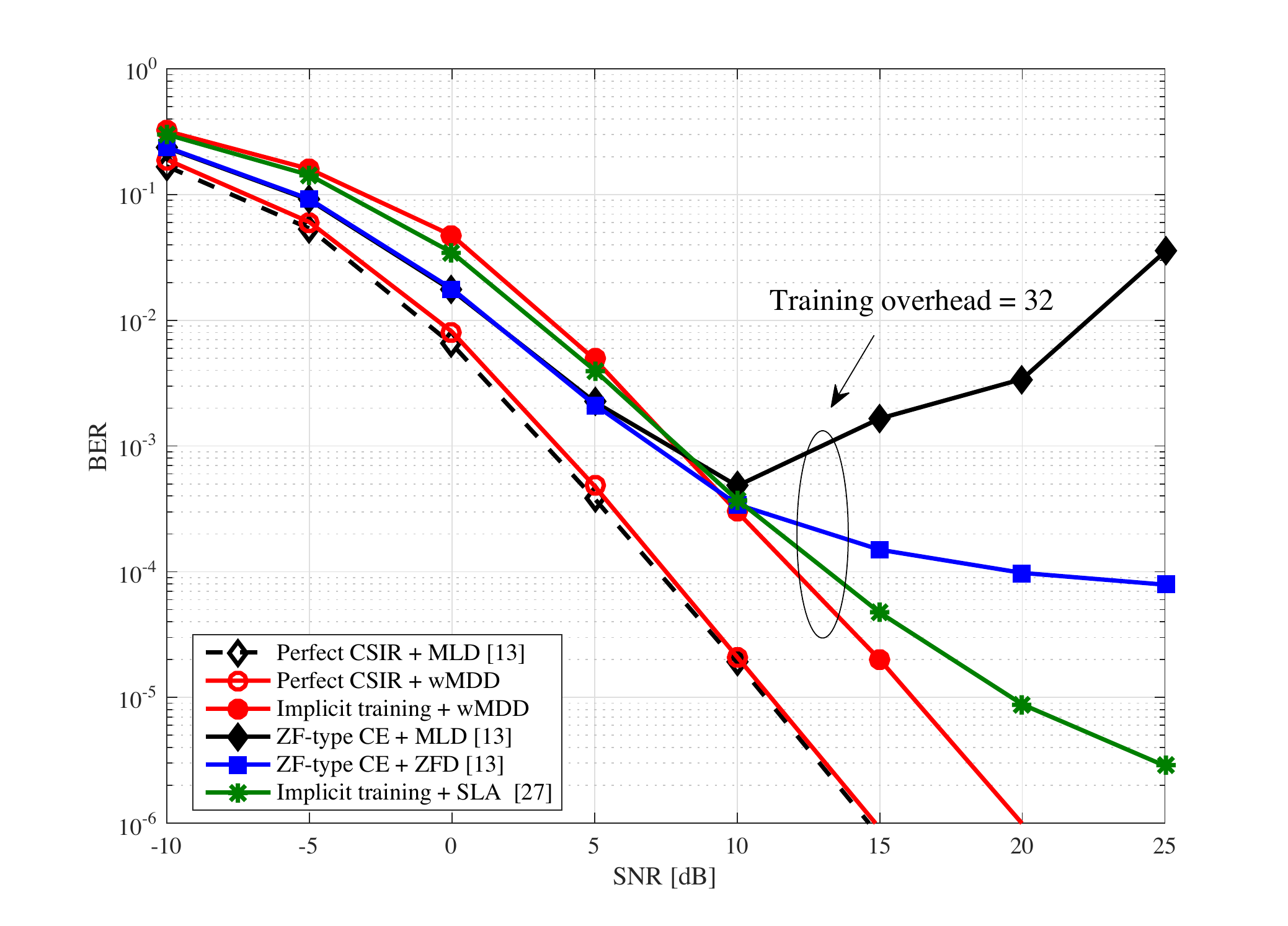}}\vspace{-0.5cm}
\caption{Performance comparisons of the various detection methods for a small-scale uplink MIMO system when QPSK modulation and one-bit ADCs are employed with $K=2$ and $N_{\rm r}=16$. The training overhead is set to $T_{\rm o}=32$; this setting corresponds to $T=4$ for the implicit method for wMDD decoding.}
\label{comparison1}
\end{figure}

\begin{figure}
\centerline{\includegraphics[width=9.5cm]{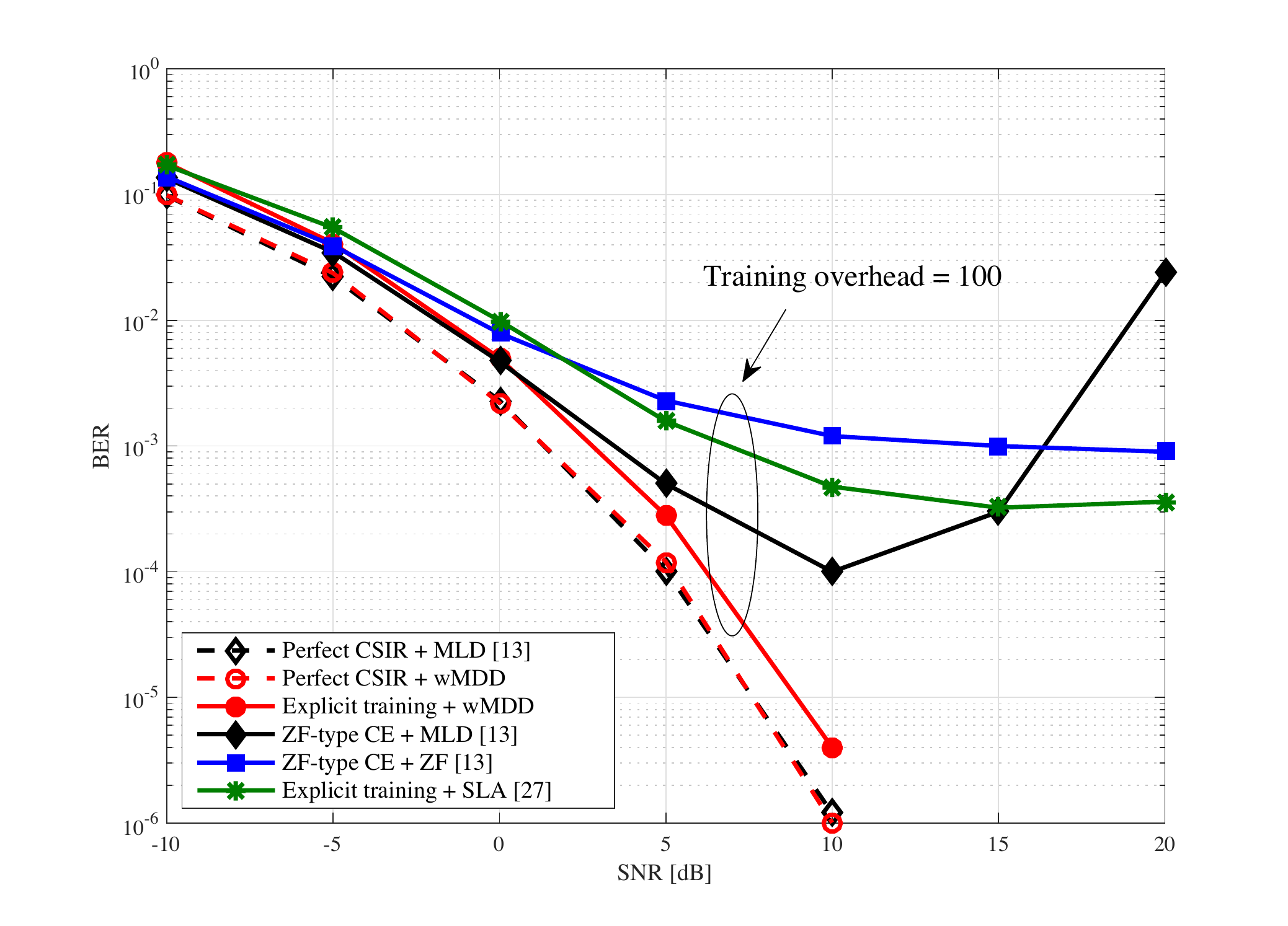}}\vspace{-0.5cm}
\caption{Performance comparisons of the various detection methods for a large-scale uplink MIMO system when QPSK modulation and one-bit ADCs are employed with $K=5$ and $N_{\rm r}=32$. The training overhead is set to $T_{\rm o}=100$.}
\label{comparison2}
\end{figure}


Fig.~\ref{comparison1} shows the BER performances of the proposed wMDD with implicit channel training method and the existing techniques. We observe that the proposed wMDD with perfect CSIR achieves the optimal MLD performance. This implies that the proposed transformation in Section~\ref{sec:Main}, from a non-linear MIMO channel to effective parallel channels, does not degrade the performance. It is remarkable that MLD with imperfect CSIR severely suffers from BER degradation especially in the high-SNR regime, due to the impact of the inaccurate CSIR. In contrast, the proposed wMDD with imperfect CSIR yields an satisfactory performance, which can outperform the existing techniques and the performance gaps grow as SNR increases. We want to emphasize that wMDD is more robust to imperfect CSIR than MLD, although both methods can achieve the optimal performance with perfect CSIR.


In Fig.~\ref{comparison2}, we consider a large-scale MIMO system for the comparisons of all the detection techniques. We can observe a similar performance trend with the case of a small-scale MIMO system in Fig.~\ref{comparison1}. An interesting observation is that the performance gap between wMDD and SLA is larger than the small-scale case. As we understood, this gap is due to the use of different distance metrics. To be specific, wMDD and SLA use the weight Hamming distance and euclidean distance, respectively, and the former seems to be more suitable for binary sequences (outputs of one-bit ADCs) than the latter.

\begin{figure}
\centerline{\includegraphics[width=9.5cm]{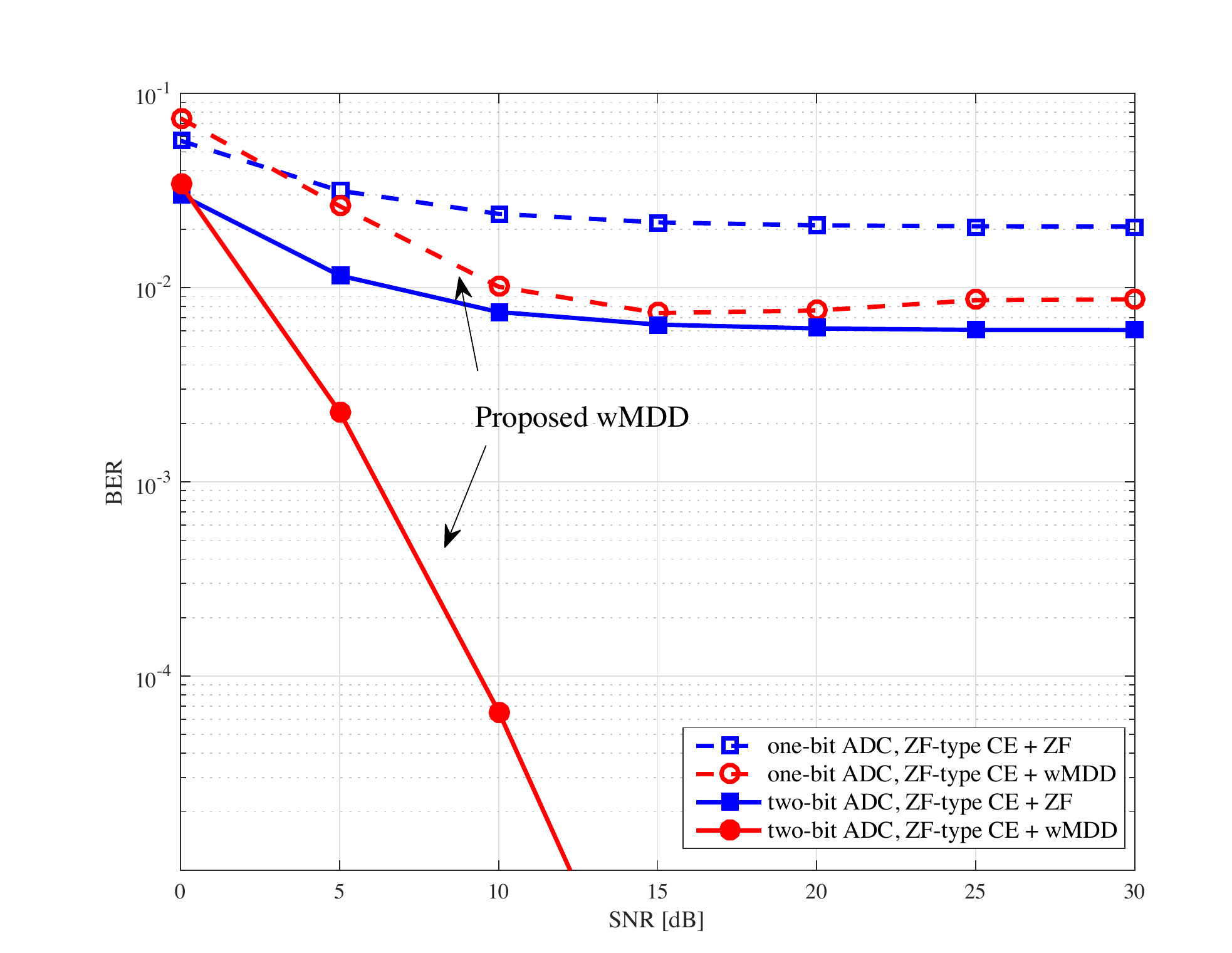}}\vspace{-0.5cm}
\caption{The BER performances for a MIMO system when 16-QAM modulation are employed with $K=4$ and $N_{\rm r}=64$. The training overhead is set to $T_{\rm o}=100$.}
\label{high-order}
\end{figure}

Finally, we see the performances of the proposed wMDD when a higher-order modulation is used. From Fig.~\ref{high-order}, we can see that both wMDD and ZFD suffer from a severe error-floor. For the wMDD, this error-floor phenomenon can be clearly explained as follows. On using 16-QAM, the number of codewords of the $\Cc$ is considerably increased compared to using QPSK, which definitely decreases the minimum distance of the $\Cc$. As it is widely known, the short minimum distance can result in an error-floor \cite{MacWilliams}. We can improve the minimum distance by increasing either code length (i.e., number of receiver antennas) or alphabet size (i.e., multi-bit ADCs). As expected, Fig.~\ref{high-order} shows that the use of two-bit ADCs significantly enhances the performances of the wMDD, by essentially increasing the minimum distance of the quaternary code $\Cc$. Most importantly, we can see that the performance of the wMDD is mainly determined by the minimum distance of the $\Cc$, rather than 
modulation order, number of users, number of receiver antennas, and ADC levels. For the ZFD, whereas, the use of two-bit ADC does not improve the slope of the BER curve, thus still suffering from the error-floor.

\section{Conclusion}\label{sec:conclusion}

Using a coding-theoretic framework, we presented a novel multiuser MIMO detection method for an uplink massive MIMO system with low-resolution ADCs. One major implication is that for MIMO systems with low-resolution ADCs, the {\em goodness} of a channel matrix $\Hm$ can be evaluated by the minimum distance of its associated code $\Cc$ as a condition number for conventional MIMO systems. We further identified that the use of unequal channel reliabilities in decoding can yield an unbounded gain. Motivated by this, we developed a novel weighted minimum distance decoding (wMDD) which can outperform the conventional minimum distance decoding (MDD) essentially with the aid of proper weights. We further considered practical communication scenarios where a BS has no knowledge of channel state information and estimates it using training sequences. In this system, we developed the communication framework based on the proposed wMDD and showed that it yields better performances than the existing detection techniques.





\begin{thebibliography}{1}

 

\bibitem{Song_Kim_Lee17} S-N. Hong, S. Kim, and N. Lee, ``Uplink multiuser massive MIMO systems with low-resolution ADCs: a coding-theoretic viewpoint," {\em Proc.  IEEE Wireless Communications and Networking Conference (WCNC),} March. 2017.

\bibitem{Larsson} E. G. Larsson, F. Tufvesson, O. Edfors, and T. L. Marzetta, ``Massive MIMO for next generation wireless systems," {\em IEEE Commun. Mag.,} vol. 52, no. 2, pp. 186-195, Feb. 2014.

\bibitem{Lu} L. Lu, G. Y. Li, A. L. Swindlehurst, A. Ashikhmin, and R. Zhang, ``An Overview of Massive MIMO: Benefits and Challenges," {\em IEEE J. Sel. Topics Sig. Process.,} vol. 8, no. 5, pp. 742-758, Oct. 2014.

\bibitem{Marzetta} T. L. Marzetta, ``Noncooperative cellular wireless with unlimited numbers of base station antennas," {\em IEEE Trans. Wireless Commun.,} vol. 9, no. 11, pp. 3590-3600, Nov. 2010.


\bibitem{Yang} H.Yang and T. L. Marzetta,``Total energy efficiency of cellular large scale antenna system multiple access mobile networks," in {\em Proc. IEEE Online Conf. Green Commun.,} Piscataway, NJ, pp. 27-32, Oct. 2013.


\bibitem{Murmann} B. Murmann, ``ADC Performance Survey 1997-2015," [Online]. Avail-able: http://web.stanford.edu/ murmann/adcsurvey.html.


\bibitem{Mezghani-2011} A. Mezghani and J. A. Nossek, ``Modeling and minimization of transceiver power consumption in wireless networks," in {\em Proc. IEEE/ITG WSA,} pp. 1-8, Feb. 2011.







\bibitem{Donnell} I. D. O'Donnell and R. W. Brodersen, ``An ultra-wideband transceiver architecture for low power, low rate, wireless systems," {\em IEEE Trans. Veh. Technology,} vol. 54, no. 5, pp. 1623-1631, Sept.  2005.

\bibitem{Hoyos} S. Hoyos, B. M. Sadler and G. R. Arce, ``Monobit digital receivers for ultrawideband communications," {\em IEEE Trans. Wireless Commun.,} vol. 4, no. 4, pp. 1337-1344, Jul. 2005.












 


%





\bibitem{Mo2} J. Mo, P. Schniter, N. G. Prelcic and R. W. Heath Jr., ``Channel estimation in millimeter wave MIMO systems with one-bit quantization," in {\em Proc. Asilomar Conference on Signals, Systems and Computers,}  pp. 957-961, Nov. 2014.


\bibitem{Risi} C. Risi, D. Persson and E. G. Larsson, ``Channel estimation and performance analysis of one-bit massive MIMO systems," [Online]. Available:http://arxiv.org/abs/1404.7736, Apr. 2014.


\bibitem{Jacobsson2015} S. Jacobsson, G. Durisi, M. Coldrey, U. Gustavsson, and C. Studer, ``One-bit massive MIMO: Channel estimation and high-order modulations," in {\em Proc. IEEE Int. Conf. Commun. (ICC) Workshop,} 2015.


\bibitem{Jacobsson2016} S. Jacobsson, G. Durisi, M. Coldrey, U. Gustavsson, and C. Studer, ``Throughput analysis of massive MIMO uplink with low-resolution ADCs," [Online] arXiv:1602.01139v2, 2016.








  





















\bibitem{Choi} J. Choi, J. Mo and R. W. Heath Jr., ``Near maximum-likelihood detector and channel estimator for uplink multiuser massive MIMO systems with one-bit ADCs," {\em IEEE Trans. Commun.,} vol. 64, no. 5, pp. 2005-2018, May 2016.


\bibitem{Li} Y. Li, C. Tao, G. Seco-Granados, A. Mezghani, A. L. Swindlehurst, and L. Liu, "Channel estimation and performance analysis of one-bit massive MIMO systems," [Online]. Available: http://arxiv.org/abs/1609.07427, Sep. 2016.

\bibitem{Mollen} C. Moll'en, J. Choi, E. G. Larsson, and R. W. Heath, Jr., ``One-bit ADCs in wideband massive MIMO systems with OFDM transmission," in {\em Proc. IEEE Int. Conf. Acoust. Speech Signal Process. (ICASSP),} Mar. 2016.


\bibitem{Mollen2} C. Moll'en, J. Choi, E. G. Larsson, and R. W. Heath, Jr., ``Uplink performance of wideband massive MIMO with one-bit ADCs,"  {\em IEEE Trans. Wireless Commun.,} vol. 16, no. 1, pp. 2156-2168, Jan. 2017.


 \bibitem{Wang2015}  S. Wang, Y. Li, and J. Wang, ``Multiuser detection in massive spatial modulation MIMO with low-resolution ADCs," {IEEE Trans.  Wireless Commun.,} vol. 14, no. 4, pp. 2156-2168, April 2015.




\bibitem{Liang} N. Liang and W. Zhang, ``Mixed-ADC massive MIMO," {\em IEEE J. Sel. Areas Commun.,} vol. 34, no. 4, pp. 983-997, Apr. 2016.

\bibitem{Wen} C.-K. Wen, C.-J. Wang, S. Jin, K.-K. Wong, and P. Ting, ``Bayes-optimal joint channel-and-data estimation for massive MIMO with low-precision ADCs," {\em IEEE Trans. Signal Process.,} vol. 64, no. 10, pp. 2541-2556, May 2016.

\bibitem{Studer} C. Studer and G. Durisi, ``Quantized massive MU-MIMO-OFDM uplink," {\em IEEE Trans. Commun.,} vol. 64, no. 6, pp. 2387-2399, 2016.




\bibitem{Choi2} J. Choi, D. J. Love, D. R. Brown III, and M. Boutin, ``Quantized distributed reception for MIMO wireless systems using spatial multiplexing," {\em IEEE Trans. Signal Process.,} vol. 63, no. 13, pp. 3537-3548, July 2015.

\bibitem{Mezghani2} A. Mezghani, M.-S. Khoufi, and J. A. Nossek, `` A modified MMSE receiver for quantized MIMO systems," in {\em Proc. Int. ITG Workshop Smart Antennas (WSA),} Mar. 2007.


\bibitem{MacWilliams} MacWilliams, F. Jessie and N. J. A. Sloane, {\em The theory of error correcting codes,} Elsevier, 1977.

\bibitem{Lok} T. M. Lok and V.K.-W. Wei, ``Channel estimation with quantized observations," in {\em Proc. IEEE Int. Symp. Inf. Theory (ISIT),} Cambridge, MA, pp. 333, Aug. 1998.

\bibitem{Ivrlac} M. T. Ivrlac and J. A. Nossek, ``On MIMO channel estimation with single-bit quantization," in {Proc. Int. ITG Workshop on Smart Antennas (WSA),} Vienna, Austria, Feb. 2007.

\bibitem{Zymnis} A. Zymnis, S. Boyd and E. Candes, ``Compressed sensing with quantized measurements," {\em IEEE Signal Process. Lett.,} vol. 17, no. 2, pp. 149-152, Feb. 2010.


\bibitem{Jeon_Hong_Lee} Y-S. Jeon, S-N. Hong, and N. Lee, ``Supervised-Learning-Aided Communication Framework for Massive MIMO systems With Low-Resolution ADCs," {\em Submitted to IEEE Trans. Signal Processing,} Mar. 2017.
 

















\end{thebibliography}
\end{document}